\documentclass[useAMS,usenatbib,usegraphicx]{mn2e}
\usepackage{amsmath,amssymb}
\usepackage{graphicx,mathptm}
\usepackage{times}
\usepackage{natbib}
\usepackage{psfig}

\newcommand{\MBH}{M_{\rm BH}}
\newcommand{\MBulge}{M_{\rm Bulge}}
\newcommand{\fedd}{f_{\rm Edd}}
\newcommand{\Lbol}{L_{\rm bol}}
\newcommand{\Ledd}{L_{\rm Edd}}

\title[The clustering of quasars in semi-analytic models] {Modeling the
  cosmological co-evolution of supermassive black holes and galaxies: II. The
  clustering of quasars and their dark environment}

\author[Bonoli et al.]
{
Silvia Bonoli$^1$\thanks{E-mail:bonoli@mpa-garching.mpg.de},
Federico Marulli$^2$,
Volker Springel$^1$, 
Simon D.M. White$^1$, 
\newauthor
Enzo Branchini$^3$,
Lauro Moscardini$^{2,4}$
\\ 
$^1$Max-Planck-Institut f\"ur Astrophysik, Karl-Schwarzschild Strasse 1,
D-85740 Garching, Germany\\
$^2$Dipartimento di Astronomia, Universit\`a degli Studi di Bologna, 
via Ranzani 1, I-40127 Bologna, Italy \\
$^3$Dipartimento di Fisica, Universit\`a 
degli Studi ``Roma Tre'', via della Vasca Navale 84, I-00146 Roma, Italy\\
$^4$INFN, Sezione di Bologna, viale Berti Pichat 6/2, I-40127 Bologna, Italy\\
}

\begin{document}



\maketitle

\label{firstpage}

\begin{abstract}
  We use semi-analytic modeling on top of the Millennium simulation to study
  the joint formation of galaxies and their embedded supermassive black holes.
  Our goal is to test scenarios in which black hole accretion and quasar
  activity are triggered by galaxy mergers, and to constrain different models
  for the lightcurves associated with individual quasar events. In the present
  work we focus on studying the spatial distribution of simulated quasars.  At
  all luminosities, we find that the simulated quasar two-point correlation
  function is fit well by a single power-law in the range $0.5 \la r \la 20\,
  h^{-1} \rm{Mpc}$, but its normalization is a strong function of
  redshift.  When we select only quasars with luminosities within the range
  typically accessible by today's quasar surveys, their clustering strength
  depends only weakly on luminosity, in agreement with observations.  This
  holds independently of the assumed lightcurve model, since bright quasars are
  black holes accreting close to the Eddington limit, and are hosted by dark
  matter haloes with a narrow mass range of a few $10^{12} h^{-1}
  \rm{M}_{\odot}$. 
  Therefore the clustering of bright quasars cannot be used to disentangle
    lightcurve models, but such a discrimination would become possible
  if the observational samples can be pushed to significantly fainter limits.
  Overall, our clustering results for the simulated quasar population agree
  rather well with observations, lending support to the
  conjecture that galaxy mergers could be the main physical process
  responsible for triggering black hole accretion and quasar activity.
\end{abstract}


\section{Introduction}

At the beginning of this century, very bright quasars powered by black holes
(BHs) with masses of the order of $10^{9} \rm{M}_{\odot}$ were discovered at
redshifts up to $z \sim 6$ \citep{fan00,fan01}.  At the same time, X-ray
observations showed that the space density of Active Galactic Nuclei (AGN)
peaks at $z \sim 2-3$, and AGN with high X-ray luminosities are more common at
higher redshift with respect to their low-luminosity counterpart
\citep{steffen03, cowie03, cattaneo03, ueda03, hasinger05}.  \citet{heckman04}, using
optical data from SDSS, found that at low redshift only BHs with a mass $\la
10^{7} \rm{M}_{\odot}$ are actively growing.  Combined, these observations
suggest that supermassive black holes (SMBHs) grow ``anti-hierarchically'': the more massive BHs were
already in place at high redshift, and since then the accretion activity has
shifted to smaller scales.

Understanding how this evolution of BH growth relates to cosmic structure
formation, how BH accretion depends on the environment, and how BHs interact
with their host galaxies, have become central questions of cosmology that need
to be answered for a full understanding of galaxy formation. In fact, it has
become clear in recent years that BHs and galaxies are linked and mutually
influence each other. This co-evolution has been explored in recent years
through analytic, semi-analytic and fully numerical approaches in numerous
studies \citep[e.g.,][]{silk98, kauffmann00, merloni04, diMatteo05, cattaneo05,
croton06a, monaco07, malbon07, marulli06, marulli07}.

Starting from \citet{soltan82}, many papers have combined the present-day BH
mass function with the AGN luminosity density of quasars at various redshifts
to conclude that most of the mass in todays BHs must have been accumulated
during phases of bright AGN activity \citep[see
also][]{yu02,elvis02,marconi04,merloni08}. The duration of these
highly-efficient accretion phases could range from a few $10^{7} \rm{yr}$
\citep{yu02} up to $10^{8} \rm{yr}$ \citep{marconi04}, values that strongly
depend on the BH mass range considered and on the assumed radiation efficiency
$\epsilon$. In fact, the precise value of this quasar lifetime is still an
open question \citep{martini04}.  Estimates of the duration of individual
accretion events using, for example, the {\em proximity effect}
\citep{carswell82,bajtlik88}, have suggested lifetimes of the order of $1 ~
\rm{Myr}$ \citep{kirkman08}.

\citet{haiman01} and \citet{martini01} suggested to use quasar clustering to
obtain estimates of the quasar lifetime \citep[see also the seminal work of
][]{cole89}. The reasoning behind this conjecture
is simple: if quasars are strongly clustered, their hosts must be rare objects,
and therefore they must also be long events in order to account for the total
quasar luminosity density observed. If, on the other hand, their clustering is
comparable to the clustering of small dark matter haloes, their hosts must be
much more common, and their luminous phases must therefore have short duration.

The advent of wide-field surveys like SDSS and 2dF quasi-stellar object (2dFQSO)
 \citep{york00,croom04}
with their observation of thousands of quasars has allowed a detailed
investigation of the clustering properties of accreting BHs. \citet{croom05}
and \citet{porciani04} calculated the correlation function of quasars observed
in 2dF in the redshift range $0.5 \la z \la 2$.  Both groups found that the
clustering strength is an increasing function of
redshift, but that it does not depend significantly on quasar luminosity.  The
inferred values of the bias would suggest that quasars of the observed
luminosities are hosted by haloes of a few $10^{12} h^{-1} \rm{M}_{\odot}$,
which remains approximately constant with redshift, since
haloes of a fixed mass are progressively more clustered towards higher
redshift \citep[see also][]{grazian04}.  Following the approach of
\citet{haiman01} and \citet{martini01},
the estimated quasar lifetime would be a few $10^{7} \rm{yr}$, reaching
$\sim 10^{8} \rm{yr}$ at the highest observed redshifts.  More recent studies
on larger samples and at different redshifts have confirmed these results
\citep{shen07, myers07, coil07, daAngela08,padmanabhan08, ross09}. However, the
magnitude range covered by these surveys is typically quite narrow, and this
may explain the lack of evidence for a significant dependence of
clustering on luminosity. When \citet{shen08b} analyze the clustering of the
$10 \%$ brightest objects of their sample, they find that these quasars have a
higher bias compared to the full sample.

Using hydrodynamical simulations of isolated galaxy mergers
\citep{springel05a, diMatteo05}, \citet{hopkins05} studied the luminosity
distribution of
accreting BHs, whose activity is triggered by the merger
event. \citet{hopkins05} found that the luminosity distribution of the simulated
AGN is equivalent to a highly efficient accretion phase (with very high Eddington
ratios), followed by a decaying phase where AGN spend most of their
life. During this extended period, they would appear as faint AGN, even though
they may, in fact, contain quite massive BHs.

Based on these results, \citet{lidz06} explored the dependence of quasar
clustering on luminosity, using an analytic approach to connect quasars, black
hole masses and halo masses.  In a quasar model in which the bright end of the
luminosity function is populated by BHs accreting close to their peak
luminosity, and the faint end is mainly populated by BHs accreting at low
Eddington ratio, there should be no strong dependence of clustering on quasar
luminosity, i.e., bright and faint AGN should actually be the same type of
objects, but seen in different stages of their evolution. They should
therefore be hosted by dark matter haloes of similar masses and hence exhibit
similar clustering properties. Assuming a relation between the quasar B-band
peak luminosity and the mass of the host haloes, \citet{lidz06} tested this
prediction, and indeed found that only a narrow range of halo masses should
host active quasars, with a median characteristic mass of $M_{\rm halo} \sim
1.3 \times 10^{13} \rm{M}_{\odot}$. As pointed out by the authors, only future
surveys that will be able to observe the faint quasars in their quiescent
stage will be able to test this picture, and to rule out the alternative
hypothesis of luminosity-dependent quasar clustering.

In the present work we explore the properties of quasar clustering using a
semi-analytic model for galaxy formation and BH accretion developed on the
outputs of the Millennium Simulation \citep{springel05b}. Compared to other
theoretical work, we do not have to make assumptions about the halo
population hosting BHs nor about the relation between the halo mass and the
quasar luminosity (or BH mass), since they are a natural outcome of the
simulation of the galaxy formation process. However, we have to make
assumptions about the physics of BH accretion, and what triggers AGN activity.
In this work we are especially interested in testing the assumption that
galaxy mergers are the primary physical mechanism responsible for triggering
accretion onto massive BHs. To this end we explore the simulation predictions
for quasar clustering and the quasar luminosity function obtained with a pure
Eddington-limited lifetime model and a model that includes a low-luminosity
accretion mode as described by \citet{hopkins05} \citep[see also][hereafter
Paper I]{marulli08}.

After an introduction to our methodology and a review of some basic properties
of our simulated AGN population (Section 2), we show their clustering
properties in Section 3, where we also compare our results with the most
recent observational optical data available. In Section 4, we show the
relation between luminous BHs, quiet BHs and their host haloes. Finally, we
summarize and discuss our results in Section 5.

\section{Models for Black Hole accretion and emission}

In this Section, after a short overview of our semi-analytic model, we
describe the different phases of BH growth and emission adopted in our model.
We then review some basic properties of the simulated AGN population; further
details are given in Paper I.

\subsection{From dark matter particles to galaxies}

The semi-analytic model used in this work is run on the outputs of the
Millennium Simulation \citep{springel05b}. This is an N-body simulation which
follows the cosmological evolution of $2160^{3} \simeq 10^{10}$ dark matter
particles, each with mass $\sim 8.6 \times 10^{8} h^{-1} \rm{M}_{\odot}$, in a
periodic box of $500 h^{-1} \rm{Mpc}$ on a side. The cosmological parameters
used in the simulation are the ones of the WMAP1 \& 2dFGRS `concordance'
$\Lambda$CDM framework: $\Omega_{m} = 0.25$, $\Omega_{\Lambda} = 0.75$,
$\sigma_{8} =0.9$, Hubble parameter $h=H_{0}/100 ~ \rm{km} \rm{s}^{-1}
\rm{Mpc}^{-1} =0.73$ and primordial spectral index $n=1$ \citep{spergel03}.

The merging history trees extracted from this simulation describe the detailed
formation history of DM haloes and their subhaloes, identified with a
friends-of-friends (FOF) group-finder and an extended version of the {\small
  SUBFIND} algorithm \citep{springel01}, respectively. Using the trees as
basic input, our semi-analytic code describes the baryonic processes of galaxy
formation and allows the prediction of galaxy properties in a large
cosmological volume.

The present work is based on the galaxy formation model described by
\citet{croton06a} and \citet{delucia07}, which we extended to follow the
details of BH accretion and the lightcurves of AGN. We refer the reader to
these papers for a full description of the baryonic physics which describes
the evolution of galaxies, their stars and their gas. Below, we describe only
the prescriptions that directly relate to our study of the evolution of
SMBHs.

\subsection{Creation and accretion of BHs }

In this subsection we discuss our modeling of the physical processes
responsible for BH accretion. In the semi-analytic model, a fraction of the
mass of a halo is assigned to baryons in the form of hot gas, which as time
evolves, will cool and form a galaxy. We also add a `seed' BH of very small
mass to each newly formed halo. As galaxies evolve their central BHs are
allowed to grow through mergers with other BHs and through gas accretion
during the `radio mode' and during the `quasar mode'. The quasar mode is the
phase during which BHs accrete most of their mass, and during which BHs can
shine as bright AGN.  We will therefore mainly concentrate most of the
discussion on this phase, and we will describe which physical process might be
responsible for triggering this activity.

\subsubsection{BH seeding}

The origin of primordial massive BHs is still subject of intense debate: SMBHs
seeds could grow out of the remnants of Pop III stars
\citep[e.g.,][]{madau01,heger02} or could have their origin directly in the
collapse of a low-angular momentum gas cloud
\citep[e.g.,][]{loeb94,koushiappas04}.  In the first case the progenitors of
SMBHs would have a mass of the order of $10^2-10^3\,{\rm M}_{\odot}$, much
less than what could be the outcome of low-angular momentum gas collapse
($10^5 {\rm M}_{\odot}$).

Unfortunately, due to exponential growth during accretion, it is very difficult
to use the
local population of massive holes to recover information about their original
mass before the onset of accretion. On the theoretical side, simulations are
being carried out to investigate which model for massive BH formation is most
plausible \citep[e.g.,][]{bromm03, alvarez08}.  Observationally, these models
for primordial BHs will hopefully be tested in the near future
either directly through gravitational wave detection \citep{sesana05,
  koushiappas06}, or indirectly by looking at the effect that primordial BHs
might have on reionization \citep[e.g.,][]{ricotti05,ripamonti08}.

As in Paper I, we assume here that every newly-formed galaxy hosts a central
BH of $10^3\, {\rm M}_{\odot}$. This seed BH may then start accreting through
the processes described below. Note however that a much larger seed would only
influence the BH evolution in our model at very high redshifts, but it would
not influence the results in the redshift range of main interest in this
paper, simply because the large growth factor 
soon cancels any information about the seed mass.

\subsubsection{Radio mode}

When a static hot halo has formed around a galaxy, we assume that a fraction
of the hot gas continues to accrete onto the central BH, causing low-level
`radio' activity in the galaxy center. For clarity, this phase, which is
called in jargon {\em radio mode} because it is associated with the activity
of radio galaxies at the centre of galaxy clusters \citep{best05}, does not
include the powerful emission of FRII radio loud QSOs.  Following
\citet{croton06a}, the BH mass accretion rate during these phases of {\em radio
mode} activity is postulated to scale as follows:
\begin{equation}
\dot{M}_{\rm BH,R} = \kappa_{\rm{AGN}} \left(\frac{M_{\rm BH}}{10^{8}
  M_{\odot}}\right) \left(\frac{f_{\rm hot}}{0.1}\right)
  \left(\frac{V_{\rm vir}}{200\,\rm{km\,s^{-1}}}\right)^3 ~, 
\label{accretionR}
\end{equation}
where $M_{\rm BH}$ is the BH mass, $f_{\rm hot}$ is the fraction of the
total halo mass in the form of hot gas, $V_{\rm vir}$ is the virial velocity of
the halo and $\kappa_{\rm{AGN}}$ is a
free parameter set equal to $7.5\times10^{-6} \rm{M_{\odot} yr^{-1}}$ in
order to reproduce the turnover at the bright end of the galaxy
luminosity function.  Since $f_{\rm hot}$ is approximately constant for
$V_{\rm vir} \gtrsim 150\,{\rm km\,s^{-1}}$, the dependence of
$\dot{M}_{\rm BH,R}$ on this quantity has  little effect. Note that the
accretion rate given by equation (\ref{accretionR}) is typically
orders-of-magnitude below the Eddington limit. In fact, the total mass
growth of BHs in the {\em radio} relative to the {\em quasar mode} (discussed below)
is negligible \citep{croton06a}.

It is also assumed that {\em radio mode feedback} injects energy
efficiently into the surrounding medium, which can reduce or even stop the
cooling flows in halo centers. The mechanical heating generated by this
kind of BH mass accretion is parameterized as $L_{\rm BH} =
\epsilon\dot{M}_{\rm BH}c^2$, where $\epsilon = 0.1$ is the {\em accretion
  efficiency} and $c$ is the speed of light. The heating modifies the infall
rate due to cooling according to:
\begin{equation}
\dot{m}_{\rm cool}' = \dot{m}_{\rm cool} - 
\frac{L_{\rm BH}}{0.5 V_{\rm vir}^2} ~.
\label{effective_cool}
\end{equation}
For consistency we never allow $\dot{m}_{\rm cool}'$ to fall below zero.  In
this scenario, the effectiveness of radio AGN in suppressing cooling flows is
greatest at late times and for large values of the BH mass, which is required
to successfully reproduce the luminosities, colors and clustering of
low-redshift bright galaxies.

\subsubsection{Quasar mode}

This is the phase during which BHs accrete cold gas and build up most of their
final mass. This phase has recently acquired the jargon name {\em quasar mode}
because it is only through the very efficient accretion of cold gas that a BH
can shine as a bright AGN, but we stress that this phase is also meant to
include accretion of cold gas at low Eddington ratios.

The tight relation observed locally between BH mass and the host bulge
\citep[e.g.,][]{magorrian98, ferrarese00, tremaine02, marconi03} suggests that
bulges and BHs might form during the same events and/or they strongly
influence each other as they evolve.  Simulations have shown that during
mergers of gas-rich disk galaxies gas is channeled toward the nuclei of the
merging galaxies through gravitational torques \citep{barnes96}, and this
process can indeed be responsible for the formation of bulges as well as for
BH accretion \citep{springel05a, diMatteo05}.

Based on these results, and following \citet{kauffmann00}, in the present work
we assume that the quasar phase is triggered by galaxy mergers. In practice,
during merger events, we assume that the BHs hosted by the merging galaxies
instantaneously coalesce and form a single BH whose mass is the sum of the
progenitor BHs, and that this resulting BH starts accreting a fraction of the
available cold gas. In Paper I we found that we need to feed BHs more
efficiently at high redshifts in order to build massive BHs by $z=5$ without
invoking super-Eddington accretion or much more massive seed masses. In this
work we assume that the amount of cold gas accreted during each merger depends
linearly on redshift \citep{croton06b}:
\begin{equation} \label{eqn:quasar_merg_z}
\Delta M_{\rm BH,Q} = \frac{f'_{\rm BH} \ m_{\rm cold}}{1 +
  (280\,\rm{km\,s^{-1}}/V_{\rm vir})^2} (1+z_{\rm merg})  \, , 
\end{equation}
where $m_{\rm cold}$ is the total mass of cold gas in the final galaxy, $z_{\rm
merg}$ is the redshift of the merger and 
\begin{equation} 
f'_{\rm merg} = f_{\rm merg}\ (m_{\rm sat}/m_{\rm central})\, ,
\label{eqn:fBH}
\end{equation}
where $f_{\rm merg}\approx 0.02$  is a normalization parameter chosen to match the observed local
$\MBH-\MBulge$ relation and $m_{\rm sat}/m_{\rm central}$ is the mass ratio of
the merging galaxies.

\begin{figure}        
        \includegraphics[width=0.45\textwidth]{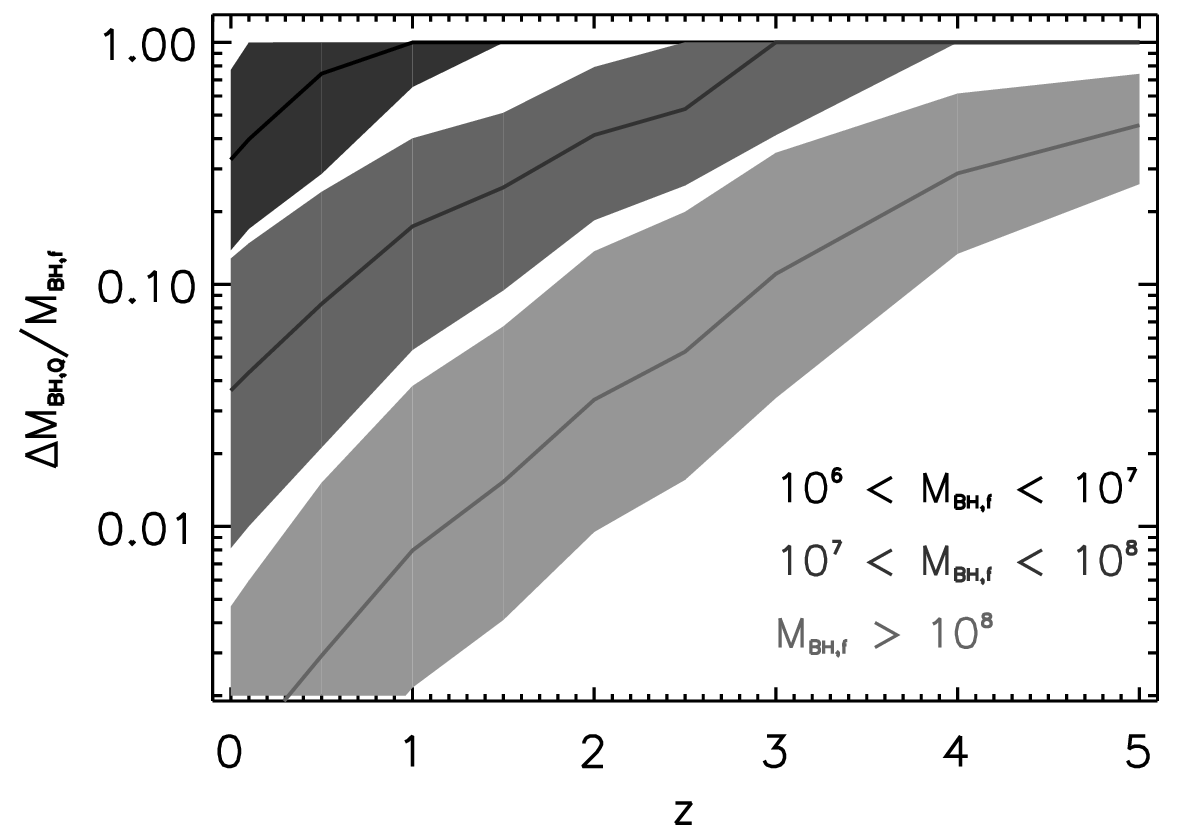}
        \caption{Median accreted gas $\Delta M_{\rm BH,Q}$ relative to the final BH mass for each
          accretion event, for three different final mass bins. The filled
          contours enclose the $25$ and $75$ percentiles.}
	\label{fig:DM_acc}
\end{figure}

In Figure \ref{fig:DM_acc} we show, as a function of redshift, the median
accreted mass $\Delta M_{\rm BH,Q}$, relative to the value of the BH mass at the end
of a single accretion event, for three final mass bins. Small-mass BHs accrete
 efficiently at all epochs (higher curve), whereas BHs that, at the end of
the accretion event, end-up in massive objects (lower
curve) accrete most of their mass at early times: at low-redshifts, the amount
of `new' gas accreted is relatively small compared to the mass already
acquired. This
behavior is in agreement with the apparent `anti-hierarchical' growth of BHs:
observations in the soft and hard X-rays have shown that the number density of
bright AGN declines with decreasing redshift, while the density of fainter
active nuclei shows the opposite trend \citep{cowie03, steffen03,
  ueda03,hasinger05}.  \citet{heckman04} used the emission lines of type 2 AGN
observed with SDSS to investigate whether the decrease of the space density of
bright objects is simply due to a decrease in the accretion rate or a decrease
in the typical mass of actively growing BHs. These authors found that the
typical mass of BHs that are today actively accreting is $\la 10^{7} ~
\rm{M}_{\odot}$, and that larger BHs are experiencing little accretion.

\begin{figure}
        \includegraphics[width=0.45\textwidth]{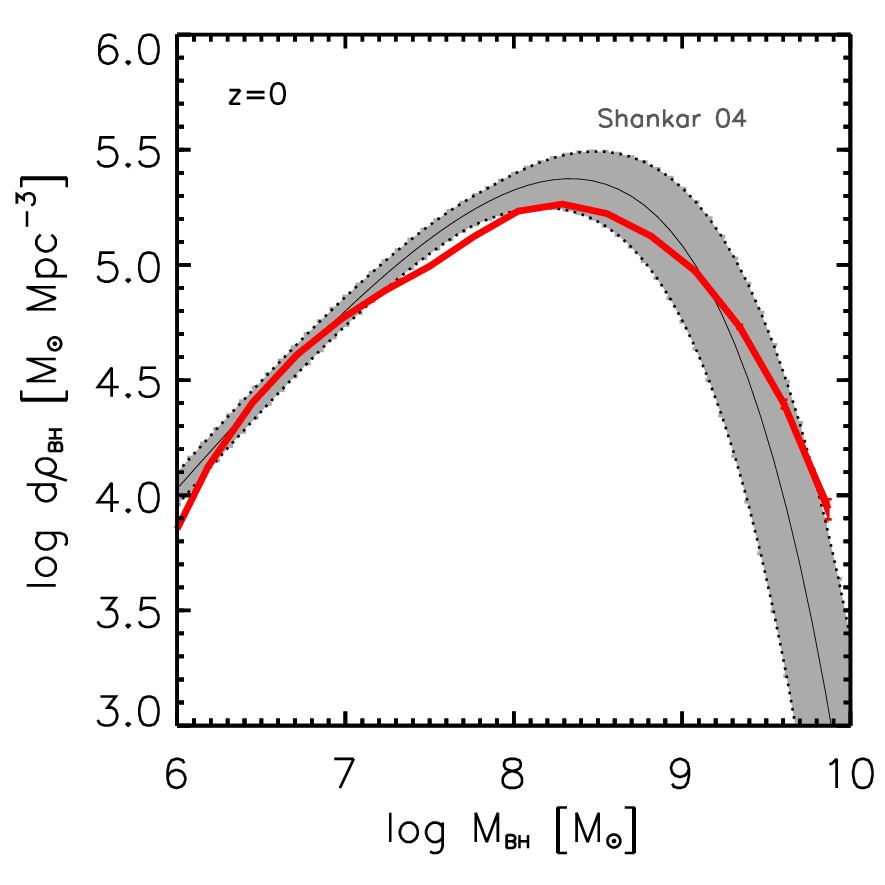}
        \caption{Differential BH mass density at $z=0$ (red thick line)
          compared to the observational estimate of \citet{shankar04} (solid
black line, with errors enclosed in the grey shaded area).}
	\label{fig:mass_func_M}
\end{figure}

In Paper I we showed that, at $z=0$, this model for BH accretion is able to
reproduce not only the observed $\MBH - \MBulge $ relation \citep{haring04},
but also other scaling relations, such as the ones between the BH mass and the
galaxy central velocity dispersion or color \citep{marconi03,ferrarese05}.
The $z=0$ differential mass density of our simulated BHs is shown in Figure
\ref{fig:mass_func_M} compared with the observational estimate of
\citet{shankar04}. The corresponding local mass density (for our cosmology
with $h =0.73$) is $\rho_{\rm BH}=3.35 \times 10^5 ~\rm{M}_{\odot} ~
\rm{Mpc}^{-3} $, which is in good agreement with \citet{graham07} (we refer to
these authors for a summary of the values quoted in the literature).

\begin{figure*}
\begin{center}
        \includegraphics[width=1.\textwidth]{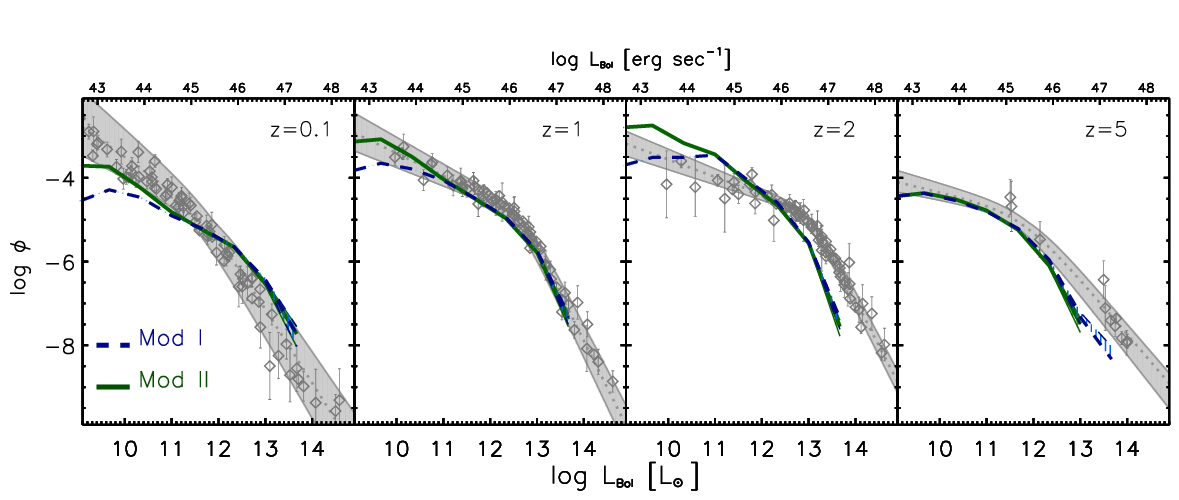}
        \caption{Bolometric luminosity function assuming Eddington-limited
          accretion (Mod I, blue-dashed curve), or Eddington-limited accretion
          followed by a quiescent phase of low luminosity (Mod II, green-solid
          curve), with errors calculated using Poisson statistics.  The
          luminosity functions are compared with the compilation of
          \citet{hopkins07a} (grey points with best fit given by the grey band).}
        \label{fig:Lum_func_ModI_ModII}
\end{center}
\end{figure*}

To study the redshift evolution of the BH population, it is important to not
only to consider the evolution of the BH mass, but also to relate this to the
radiation output of the accretion. If we are interested in the instantaneous
brightness of a quasar, we not only need to calculate how much mass it
accretes, but also how long this takes. In other words, we need to model the
lightcurve of individual phases of quasar activity. In Paper I we introduced
and tested different models for the AGN lightcurve, and we compared our
results with the AGN bolometric luminosity function of \citet{hopkins07a}. We
here briefly describe the lightcurve models adopted for the present study.

At any given time, the bolometric luminosity emitted by an accreting BH is
given by
\begin{eqnarray} \label{eq:Lagn}
\Lbol (t) & = & \epsilon\dot{M}_{\rm accr}(t)c^2=
\frac{\epsilon}{1-\epsilon}\dot{M}_{\rm BH}(t)c^2 \nonumber \\
            & = & \fedd (t) \Ledd(t)=\fedd \frac{M_{\rm BH}(t)}{t_{\rm
                Edd}}c^2 ,
\end{eqnarray}
where $\epsilon$ is the radiative efficiency, $L_{\rm Edd}$ is the Eddington
luminosity, $\fedd$ is the fraction of Eddington luminosity emitted, and
$t_{\rm Edd}=\sigma_{\rm T} c /(4\pi m_{p} G) \sim0.45\,{\rm Gyr}$ (note that 
we are here considering only the luminosity emitted during the {\em quasar mode}
phase, thus ignoring the contribution from $\dot{M}_{\rm BH,R}$).  If, at any
given time, the radiative efficiency and the Eddington ratio are known, the
accretion rate is given by:
\begin{equation} \label{eq:Mdot}
  d\ln M_{\rm BH}(t) = \frac{dt}{t_{\rm ef}(t)}, 
\end{equation}
where $t_{\rm ef}(t)=\frac{\epsilon}{1-\epsilon}\frac{t_{\rm Edd}}{\fedd(t)}$ is
the e-folding time ($t_{\rm ef}\equiv t_{\rm Salpeter}$ if $\fedd=1$).

\begin{figure}
	\includegraphics[width=0.48\textwidth]{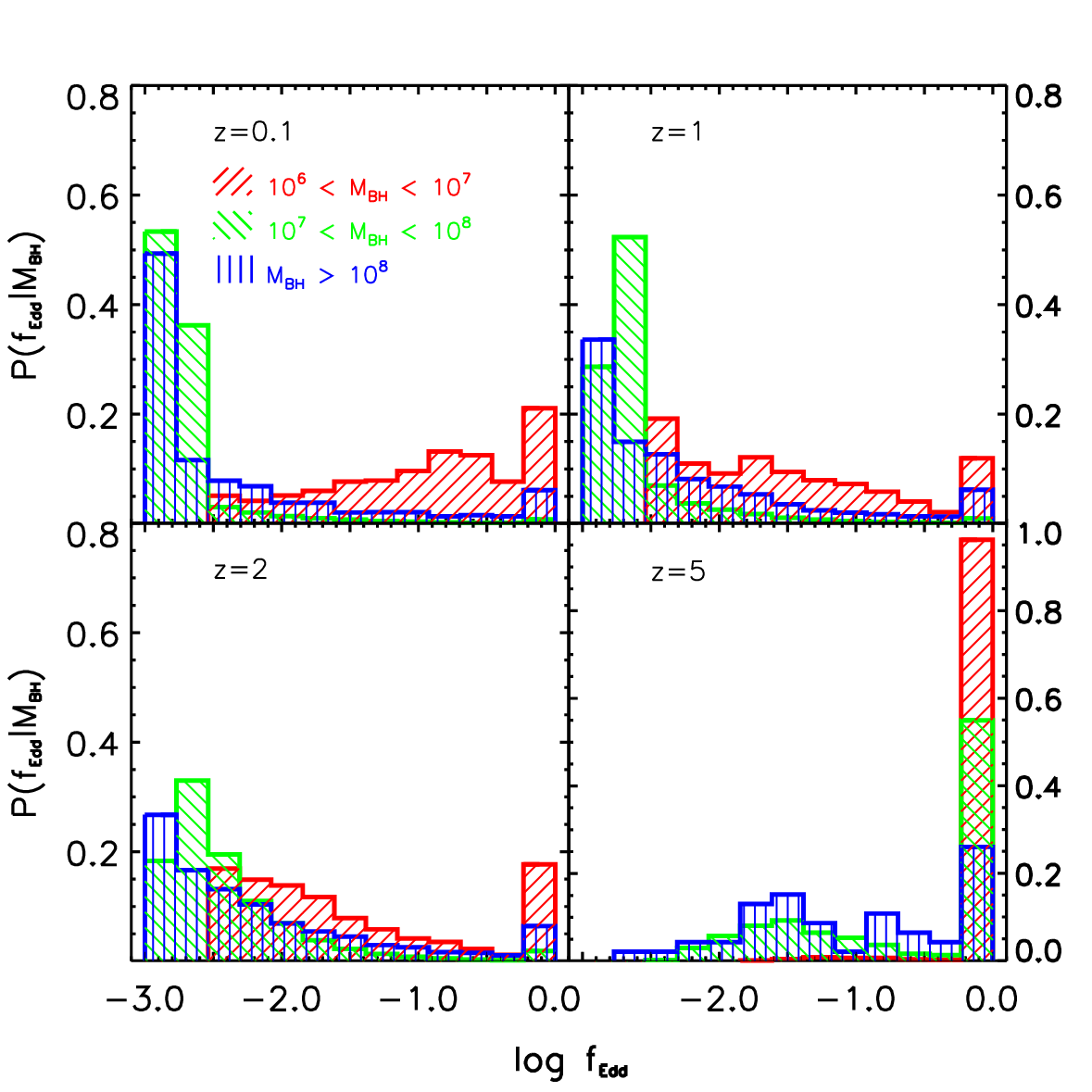} 
	\caption{Probability distribution of $f_{Edd}$, as a function of BH
          mass and redshift. The limits in the BH mass bins are shown in the
          first panel in units of $M_{\odot}$. At high redshift, most of the
          BHs accrete at the Eddington limit. Today, only the smallest BHs are
          experiencing efficient accretion.}
	\label{fig:Edd_fraction}
\end{figure}

For simplicity, we assumed a constant radiative efficiency $\epsilon =0.1$
\citep[average value for a thin accretion disk,][]{shakura73} , and we
explored different models for the time-evolution of $\fedd $. 
In this work we choose not to explore all the four models discussed in Paper I. 
Instead, we will focus on
two of them, which we regard as representative cases. The first one 
illustrates the simple case
of an AGN that shines at the Eddington luminosity. It represents a very 
simple model commonly
used in the literature, that we regard as a reference case, despite the 
fact that, as shown in Paper  I, fails to reproduce the AGN luminosity function at low and 
high redshifts.
The second model is very close to the model called 
'best' in  Paper I and
illustrates the impact of adopting a non trivial AGN light-curve, 
motivated by numerical experiments.
As discussed in Paper I this second model provides a better fit to the AGN
luminosity function.
In what follows, we present a more detailed description of the two models:

\begin{itemize}

\item {\em Model I}: $\fedd (t)= \rm{const} = 1 $. This is the simplest case,
  in which we assume that, when active, BHs accrete and radiate at the
  Eddington limit.

\item {\em Model II}: Here we assume that BHs undergo an Eddington-limited
  phase that leads to a peak luminosity $L_{\rm peak}$, which is then followed
  by a long quiescent phase at progressively lower Eddington ratios. Following
  the work of \citet{hopkins05}, we assume that in this long quiescent phase
  the average time that an AGN spends in a logarithmic luminosity interval can
  be approximated by:
     \begin{equation} \label{eqn:dt_hopkins} 
    \frac{{\rm d}t}{{\rm d}\ln \Lbol}=|\alpha|\,t_9\left(\frac{\Lbol (t)}{10^9L_\odot}\right)^\alpha,
  \end{equation}
  where $t_9\equiv t_Q(L'>10^9L_\odot)$ and $t_Q(L'>L)$ is the total AGN
  lifetime above a given luminosity $L$. \citet{hopkins05} found from merger
  simulations that $t_9\sim10^9\,{\rm yr}$ over the range $10^9L_\odot< \Lbol
  <L_{\rm peak}$; here, we assume always $t_9=10^9 \rm{yr}$.  In the range
  $10^{10}L_\odot \la L_{\rm peak}\la 10^{14}L_\odot$, \citet{hopkins05} also
  found that $\alpha$ is a function of only the AGN luminosity at the peak of
  its activity, $L_{\rm peak}$, given by $\alpha=-0.95+0.32\log(L_{\rm
    peak}/10^{12}L_\odot)$, with $\alpha=-0.2$ as an upper limit.

  In this scenario, the peak luminosity $L_{\rm peak}$ reached at the end of
  the first accretion phase is $\Ledd (M_{\rm BH,peak}) $, where
\begin{equation}\label{eqn:lum_peak}
M_{\rm BH,peak}=\MBH (t_{\rm in})+ \mathcal{F} \cdot\Delta M_{\rm
BH,Q}\cdot(1-\epsilon) .
\end{equation}
Here $\MBH(t_{\rm in})$ is the BH mass at the beginning of the accretion,
$\Delta M_{\rm BH,Q}$ is the fraction of cold gas mass accreted, and
$\mathcal{F}$ sets the fraction of gas that is accreted during the
Eddington-limited phase.  After this first phase, the BH keeps accreting the
remaining cold gas at a progressively slower rate, as described by
equation~(\ref{eqn:dt_hopkins}). In Paper I we set $\mathcal{F} = 0.7$, a
value that balances the needs of efficiently building-up massive BHs and of
explaining low-$\fedd$ BHs in the local universe. Most of the available gas is
therefore accreted during the Eddington-limited phase, and the lightcurve
  model introduced by \citet{hopkins05} is used to describe only the quiescent
phase.
\end{itemize}

A direct comparison of the luminosity functions obtained using Mod I and Mod
II is shown in Figure \ref{fig:Lum_func_ModI_ModII}.  Mod I and Mod II give a
similar population of high-luminosity AGN: bright AGN are always produced by
BHs accreting close to the Eddington limit. At high redshifts, the faint-end
of the luminosity function produced by the two models is very similar as well,
suggesting that at very high redshifts BHs of all masses typically accrete at
$\fedd =1$. It is in the faint-end of the luminosity function at low redshifts
where the two models predict a different behavior for the AGN luminosity:
only Mod II (with $\mathcal{F} = 0.7$) is able to fit the low-redshift
faint-end of the luminosity function, implying that a model in which BHs
experience long, quiescent accretion phases can indeed explain the number
density of low-luminosity AGN at low redshift. This is because in Mod II
the average lifetime of AGN is much higher (a larger fraction of time is spent
at low luminosities); it is therefore more probable to observe, at a given
redshift, an AGN shining at low luminosities. For a more detailed discussion on
this, we refer again the reader to Paper I.

We have already mentioned that observations indicate that the more massive BHs
have accreted most of their mass at early times, whereas in the local universe
BHs with a mass $\la 10^{7} ~ \rm{M}_{\odot}$ are accreting efficiently
\citep{heckman04}.  These results have been confirmed more recently by
\citet{netzer07}, who found that at all redshifts $\fedd$ is smaller for
larger mass BHs. Similar compilations that use emission lines to estimate
Eddington ratios have shown that the $\fedd$ of quasars seems to be
log-normally distributed, with a peak around $\fedd \approx 10^{-1}-10^{-0.6}$
\citep{kollmeier06, shen08a}.  In Figure \ref{fig:Edd_fraction} we show, for
Mod II, the redshift evolution of the probability distribution $P(f_{\rm
Edd}|M_{\rm BH})$ of the
Eddington ratios, given the BH mass. At high
redshifts all BHs accrete close to the Eddington limit. At lower redshifts
instead only the smaller BHs are accreting at high Eddington ratios, while
the more massive ones accrete at much lower rates. Note that this figure
includes all active BHs from our simulation, and therefore a direct comparison
with observed data is not possible. We postpone a more detailed analysis of
this point to a future work, but we stress that a model with a quiescent phase
could account for the low-redshift behavior of the more massive BHs
\citep[see also the recent work of][]{hopkins08}.

\section{clustering properties}

In this section we discuss the clustering properties of our simulated AGN
sample. We first compare the predicted two-point correlation with the
autocorrelation of the DM particles.  We then compare the AGN clustering
properties with the clustering of the dark matter haloes of the Millennium
simulation, and in particular examine the differences between Mod I and Mod
II. We then explore the luminosity dependence of the clustering of the
global AGN population and of an optically-visible sub-sample. Finally, we
directly compare the clustering of our simulated $L_{\ast}$ quasars with 
recent observational results.

\begin{figure*}
\begin{center}
        \includegraphics[width=1.0\textwidth]{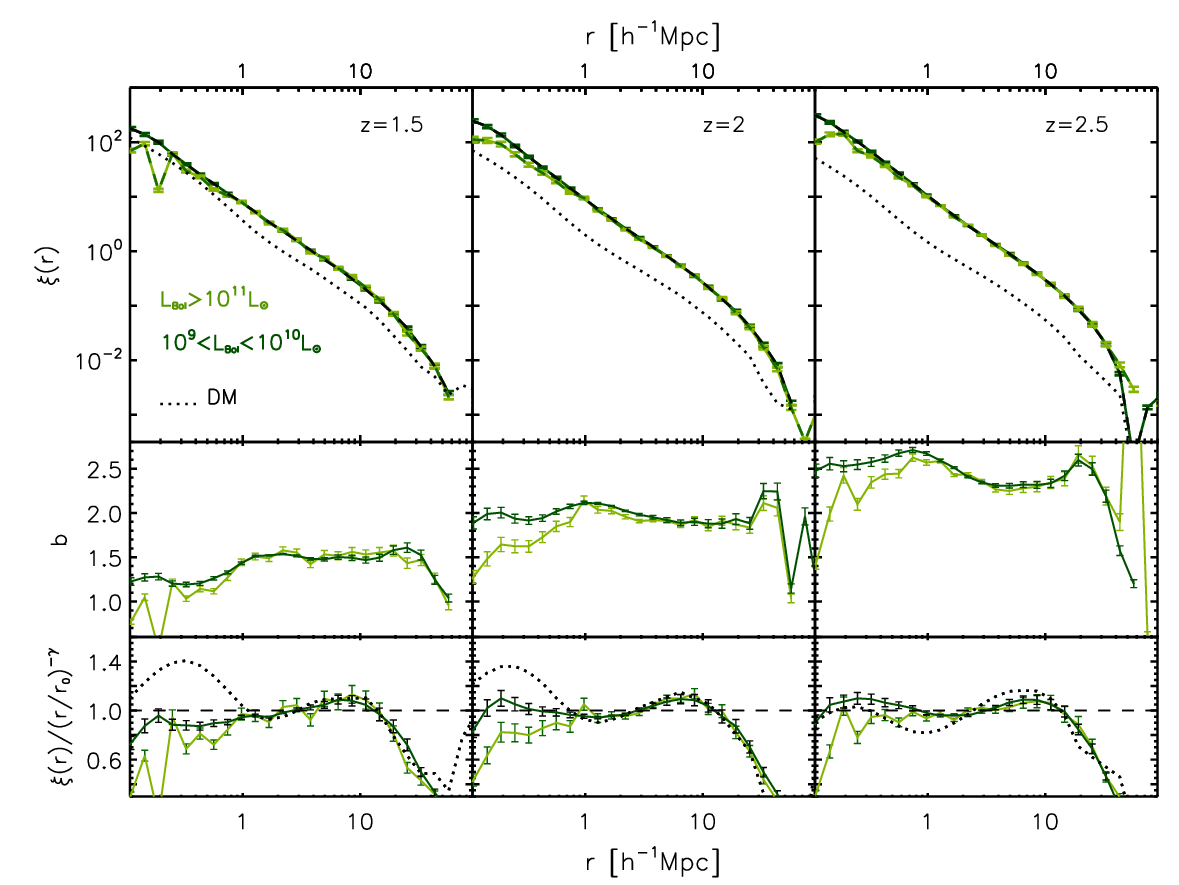}
        \caption{\textit{Upper panels}: two-point correlation function of the
          Millennium dark matter particles (dotted line) compared with the correlation of
          the AGN population, divided into a faint and a luminous sub-sample,
          depending on their bolometric luminosity (as indicated in the first
	  panel. 
	  \textit{Central panels}:
          bias between the two AGN samples and the dark matter as a function
          of scale. \textit{Lower panels}: two-point correlation from the
          upper panels divided by a power-law fit.  If $\xi(r)$ was a perfect
          power-law, the ratio should be constant with scale and equal to
          unity (dashed horizontal line). We refer to the text for a
          description of the errors.}
        \label{fig:corr_func_halo_ModI_bias}
\end{center}
\end{figure*}

\subsection{Brief description of the correlation parameters
used}\label{sec:intro_clustering}

We use the standard definition of the {\em two-point spatial correlation function} as
the excess probability for finding a pair of objects at a distance {\em r},
each in the volume element ${\rm d}V_{1}$ and ${\rm d}V_{2}$
\citep[e.g.,][]{peacock99}:
\begin{equation}\label{eqn:two-point}
{\rm d}P=n^{2} \left [ 1+ \xi(r) \right ] {\rm d}V_{1} {\rm d}V_{2},
\end{equation}
where $n$ is the average number density of the set of objects under consideration. 

The {\em clustering length} $r_{0}$ is defined as the scale at which the
two-point correlation function is unity: $\xi ({r_{0}}) \equiv 1$ (i.e., the
scale at which the probability of a pair is twice the random). At scales
between $\sim 1 ~ h^{-1} \rm{Mpc}$ up to few tens of $\rm{Mpc}$ the observed
quasar correlation function can be approximated by a power-law, usually
expressed as:
\begin{equation}\label{eqn:csi_power-law}
\xi(r)=\left( \frac{r}{r_{0}}\right)^{-\gamma}.
\end{equation}
To calculate $r_{0}$, unless otherwise stated, we will fit the two-point
correlation with such a power-law in the range $1 < r < 20 ~ h^{-1} \rm{Mpc}$
(see the next subsection for details on this).

Finally, the {\em bias} between two classes of objects (e.g., AGN and dark
matter) is defined as the square-root of the ratio of the corresponding
two-point correlation functions:
\begin{equation} \label{eqn:bias}
b_{\rm AGN,\, DM} \equiv \sqrt{\frac {\xi_{\rm AGN} (r)} {\xi_{\rm DM} (r)}}.
\end{equation}

In principle, an accurate determination of the `cosmic-variance' errors of
these quantities as measured from the simulation could be calculated from the
variance over many different realizations of the universe. As we have only one
simulation as large as the Millennium run at our disposal, this is not
practical. A reasonable alternative is to estimate the errors by subdividing
the whole Millennium volume into sub-cubes, and then by calculating the
variance among the measurements for each of these sub-volumes, an approach we
will follow here.

In order to directly estimate the impact of the cosmic
variance in the predicted AGN clustering, it is necessary to model the
AGN properties in mock samples designed to match the real ones. We have
followed this approach in a parallel work (\citet{marulli09}, submitted to
MNRAS), where
we have used the same semi-analytic model presented here to construct
mock AGN catalogues  mimicking the Chandra deep fields.

\subsection{AGN and dark matter clustering}

We here show the results for the shape of the two-point correlation of the AGN
sample, comparing it to the one of the Millennium dark matter particles. For
simplicity, we present only the results obtained with Mod II, since the
conclusions of this subsection are independent of the assumed model for the
lightcurve.

In the top panels of Figure \ref{fig:corr_func_halo_ModI_bias}, we plot the
two-point correlation of the DM particles (dotted line) and the two-point
correlation of  faint ($L_{\rm Bol}<10^{10}\rm{L}_{\odot}$) and
bright ($L_{\rm Bol}>10^{11}\rm{L}_{\odot}$) AGN (dashed lines), at three
different redshifts.  As can be seen at a glance, the main difference between
$\xi_{\rm DM}(r)$ and $\xi_{\rm AGN}(r)$ lies in the normalization, they are
substantially biased relative to each other. This bias ($[\xi_{\rm AGN}(r)
/\xi_{\rm DM}(r)]^{1/2} $) is plotted in the next set of panels of
Fig.~\ref{fig:corr_func_halo_ModI_bias}. The bias is approximately
scale-independent (at least in the range $1 < r < 20 ~ h^{-1} \rm{Mpc}$), and
its average value increases with redshift. The error $\sigma_{\log \xi_{\rm AGN}}(r)$ 
of the two-point correlation  is here the variance (in
log-space) of the two-point correlation functions calculated in eight
sub-volumes. The errors on the bias have been calculated assuming negligible
error for DM autocorrelation. By error propagation, the error on the bias is
then $\sigma_{b}(r)=  ~ b(r)  ~ \sigma_{\log \xi_{\rm AGN}}(r) (\ln 10) /2 $.

Finally, in the lower panels of Fig.~\ref{fig:corr_func_halo_ModI_bias} we
show how the two-point correlations deviate from a power-law, that is, we
divide the calculated $\xi(r)$ by the fit calculated using
eq. (\ref{eqn:csi_power-law}). As is well known \citep[e.g.][]{springel05b},
the DM correlation function deviates from a pure power-law at low and
intermediate scales. The AGN correlation function shows a similar shape at
intermediate scales ($r \sim \rm{few} ~ h^{-1} \rm{Mpc}$), but not at small scales,
where the AGN two-point correlation function is a significantly `better'
power-law that of the DM. This is highly reminiscent of the findings for the
clustering of galaxies \citep{springel05b}.

As we will again see in the next subsection, the lack of a strong correlation
signal at small scales is due to the fact that our BHs accrete gas and can
shine as bright AGN only after merger events, which, in our model,
happen mainly in the central galaxies of dark matter haloes (whose mean
separation is $\approx 1 ~ h^{-1} \rm{Mpc}$). Also note that each of our
mergers lights up only one BH, the merged BH of the two progenitor galaxies,
i.e.~our model does not account for the possibility that the two BHs exhibit
activity as a close quasar pair already prior to coalescence.  Also, as a BH is
still accreting cold gas, it can happen that its host halo merges with another
halo which could have at its center another accreting BH. This is also why the
correlation power at scales $\la 1 ~ h^{-1} \rm{Mpc}$ is non-zero, but
negligible. In a forthcoming paper we will compare a pure merger-triggered AGN
scenario, with a model in which the possible galaxy disk instability also
could contribute in feeding BHs.  In this last case we expect a larger AGN
halo occupation distribution (number of AGN in a single halo), and a different
behavior in the small-scale clustering regime.

\subsection{AGN and halo clustering}\label{sec:cluster_AGN_halo}

\begin{figure}
\begin{center}
        \includegraphics[width=0.48\textwidth]{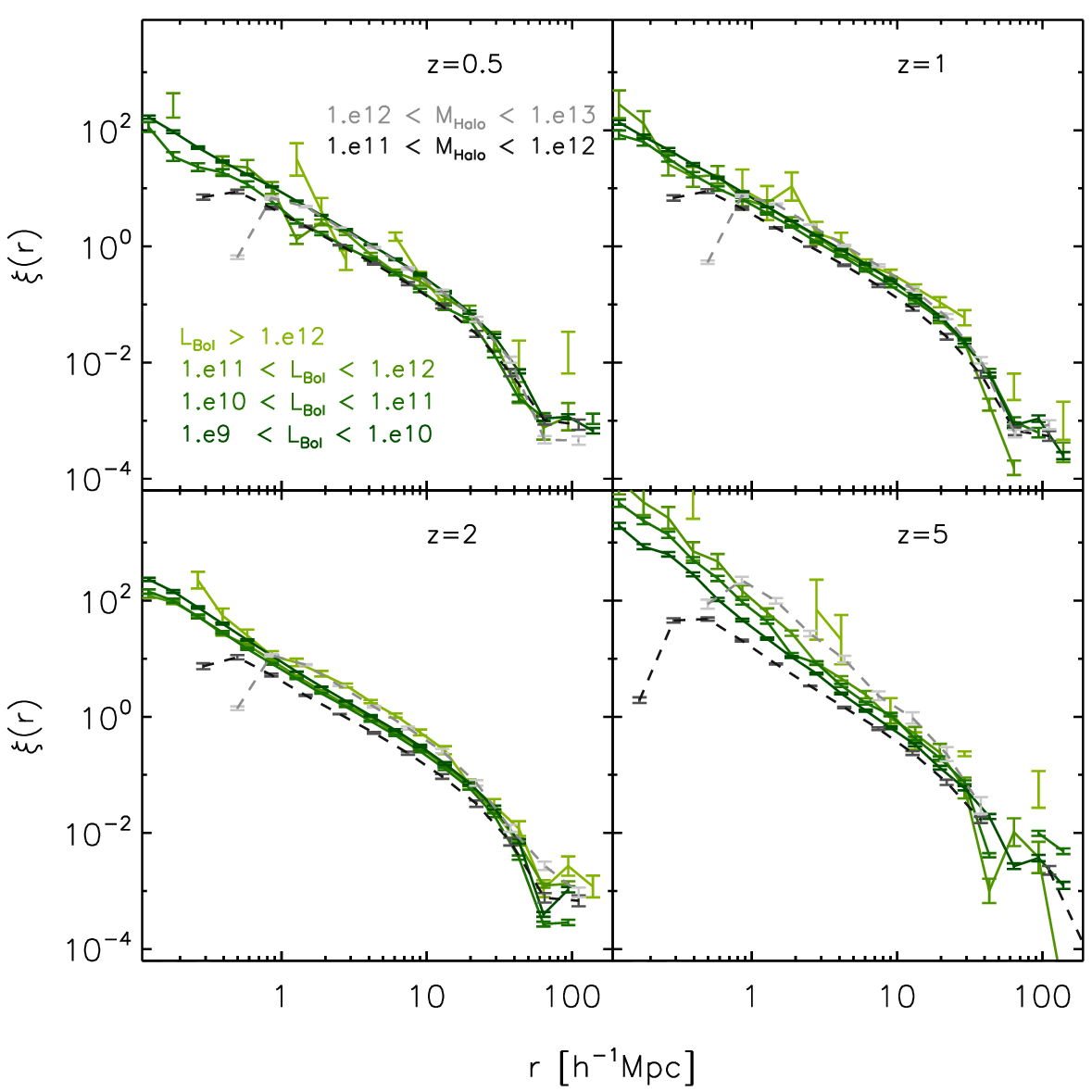}
        \caption{Two-point correlation function for the AGN sample compared to
          the two-point correlation of the Millennium FOF haloes, at various
          redshifts. The AGN are divided into 4 luminosity bins (depending on
          the bolometric luminosity), whereas the haloes are divided into two
          bins, depending on the value of their virial mass in units of
          $h^{-1} \rm{M}_{\odot}$. The AGN in this figure have been obtained
          using Mod II for the lightcurve. In Figure
          \ref{fig:corr_faintAGN_halo_ModI_ModIII}, the main difference in the
          correlation between the two models is highlighted.}
        \label{fig:corr_func_halo_ModI}
\end{center}
\end{figure}

\begin{figure}
\begin{center}
	\includegraphics[width=0.45\textwidth]{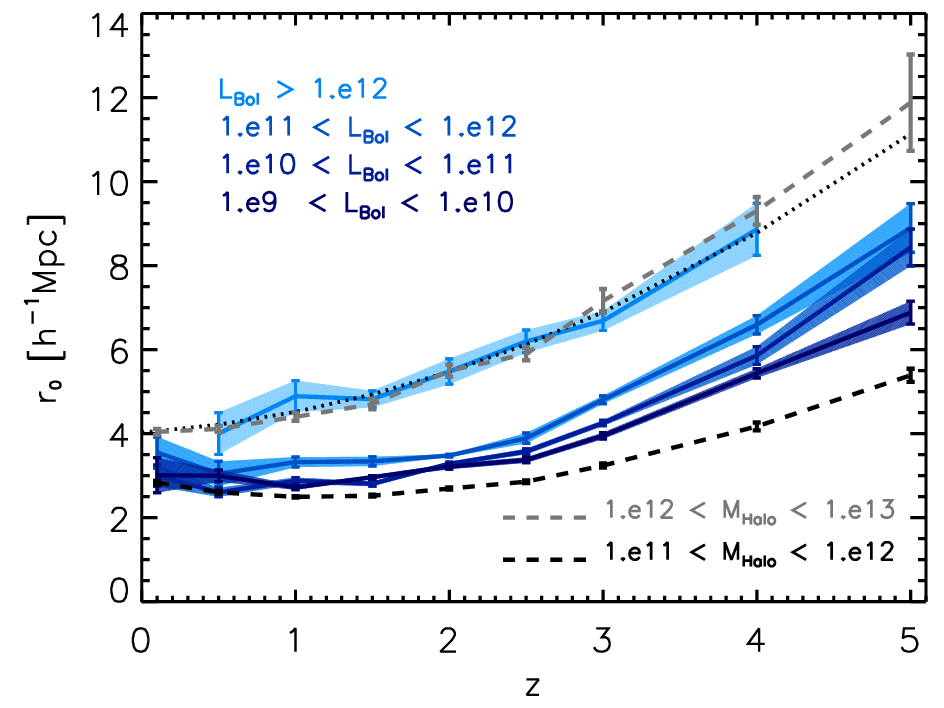}        
	\includegraphics[width=0.45\textwidth]{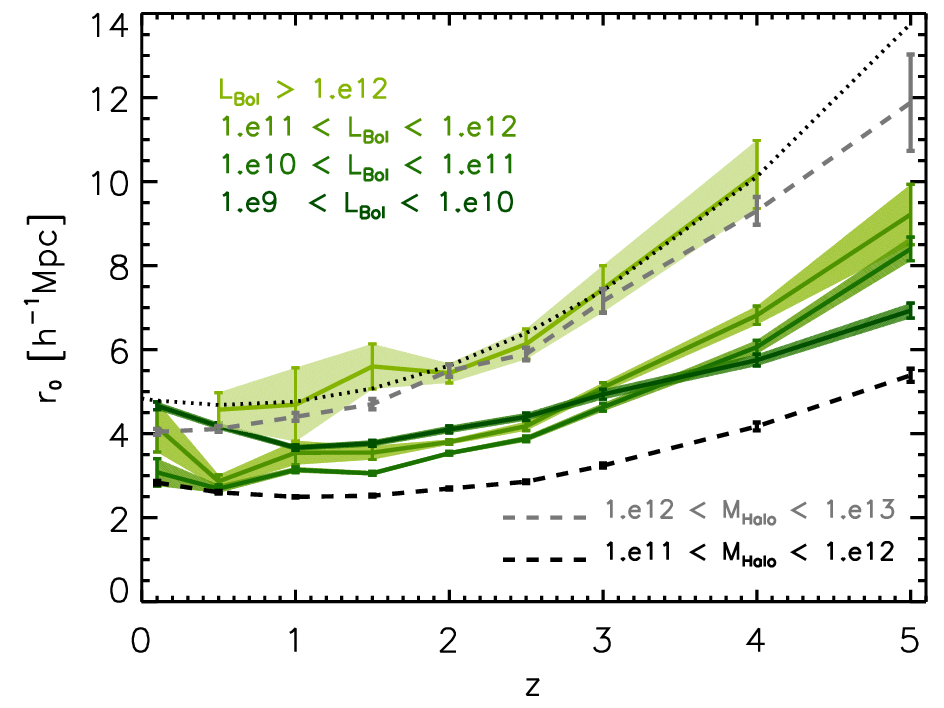}
        \caption{Correlation length as a function of redshift of the AGN
          sample divided in four bolometric luminosity bins, compared with the
          correlation length of the length of two mass bins of the Millennium
	FOF haloes. The AGN have been obtained using Mod I (upper panel) or Mod II
          (lower panel) as lightcurve models, respectively. Fits to the
          brightest bins are shown with the dotted curve.}
        \label{fig:corr_length_halo_ModI_ModIII}
\end{center}
\end{figure}

\begin{table*}
 \begin{center}
  \begin{tabular}{c|cc|cc|cc|cc|}
	 \multicolumn{9}{||c||}{Mod I}  \\
	\hline
	\hline
	\multicolumn{1}{|c|}{ }	&\multicolumn{2}{|c|}{$L_{1}$} &\multicolumn{2}{|c|}{$L_{2}$} &\multicolumn{2}{|c|}{$L_{3}$} &\multicolumn{2}{|c|}{$L_{4}$}  \\
	\hline
	z   & $r_{0}$  & $\gamma$ & $r_{0}$  & $\gamma$ & $r_{0}$  & $\gamma$  & $r_{0}$  & $\gamma$ \\
	\hline
	0.1 & - & - & $3.55\pm0.37$ & $1.4$ & $3.0\pm0.26$ & $1.79$ & $3.01\pm0.42$ & $1.5$  \\
	0.5 & $4.0\pm0.5$ & $1.69$ & $3.04\pm0.29$ & $1.49$ & $2.6\pm0.11$ & $1.53$ & $3.0\pm0.14$ & $1.49$  \\
	1.0 & $4.89\pm0.37$ & $1.62$ & $3.32\pm0.12$ & $1.63$ & $2.88\pm0.08$ & $1.5$ & $2.72\pm0.06$ & $1.52$  \\
	1.5 & $4.82\pm0.2$ & $1.79$ & $3.34\pm0.11$ & $1.57$ & $2.81\pm0.06$ & $1.56$ & $2.96\pm0.04$ & $1.57$  \\
	2.0 & $5.48\pm0.3$ & $1.71$ & $3.48\pm0.03$ & $1.55$ & $3.28\pm0.04$ & $1.55$ & $3.22\pm0.06$ & $1.5$  \\
	2.5 & $6.2\pm0.27$ & $1.54$ & $3.89\pm0.12$ & $1.58$ & $3.57\pm0.07$ & $1.54$ & $3.37\pm0.07$ & $1.55$  \\
	3.0 & $6.69\pm0.23$ & $1.79$ & $4.81\pm0.09$ & $1.6$ & $4.25\pm0.06$ & $1.59$ & $3.95\pm0.08$ & $1.57$  \\
	4.0 & $8.86\pm0.62$ & $1.77$ & $6.59\pm0.22$ & $1.76$ & $5.86\pm0.21$ & $1.7$ & $5.44\pm0.11$ & $1.66$  \\
	5.0 & - & - & $8.89\pm0.58$ & $2.04$ & $8.43\pm0.44$ & $1.89$ & $6.88\pm0.27$ & $1.81$  \\
	\hline
	\hline
	\multicolumn{9}{|c|}{fit for $L_{1}$: $r_{0}=a+b ~(1+z)+c~(1+z)^2$, with $a, b, c =[4.01, -0.21, 0.23]$}\\
	\hline
  \end{tabular}
  \caption{Values of the correlation lengths shown in the upper panel of
    Figure~\ref{fig:corr_length_halo_ModI_ModIII}. We also added the values of
    the corresponding power-law slope $\gamma$. $L_{1}$ corresponds to the
    brighter bin, $L_{4}$ to the faintest. We also give the values of the
    parameters of the quadratic fit done on $r_{0}$ for the brightest bin.}
  \label{table:r0_gamma1}
 \end{center}
\end{table*}

\begin{table*}
 \begin{center}
  \begin{tabular}{*{9}{|c|}}
	\multicolumn{9}{||c||}{Mod II}  \\
	\hline
	\hline
	\multicolumn{1}{|c|}{ }	&\multicolumn{2}{|c|}{$L_{1}$} &\multicolumn{2}{|c|}{$L_{2}$} &\multicolumn{2}{|c|}{$L_{3}$} &\multicolumn{2}{|c|}{$L_{4}$}  \\
	\hline
	z   & $r_{0}$  & $\gamma$ & $r_{0}$  & $\gamma$ & $r_{0}$  & $\gamma$  & $r_{0}$  & $\gamma$  \\
	\hline
	0.1 & - & - &  $4.15\pm0.6$ & $1.69$ & $3.08\pm0.33$ & $1.72$ & $4.66\pm0.09$ & $1.61$\\
	0.5 & $4.57\pm0.96$ & $1.96$ &  $2.86\pm0.15$ & $1.27$ & $2.69\pm0.1$ & $1.45$ & $4.18\pm0.08$ & $1.58$\\
	1.0 & $4.69\pm0.88$ & $1.62$ &  $3.55\pm0.28$ & $1.58$ & $3.14\pm0.06$ & $1.51$ & $3.67\pm0.07$ & $1.56$\\
	1.5 & $5.6\pm0.53$ & $1.89$ &  $3.55\pm0.16$ & $1.52$ & $3.05\pm0.05$ & $1.53$ & $3.77\pm0.07$ & $1.56$\\
	2.0 & $5.44\pm0.23$ & $1.68$ &  $3.8\pm0.06$ & $1.54$ & $3.53\pm0.04$ & $1.56$ & $4.1\pm0.09$ & $1.58$\\
	2.5 & $6.13\pm0.36$ & $1.52$ &  $4.18\pm0.11$ & $1.57$ & $3.88\pm0.07$ & $1.56$ & $4.4\pm0.09$ & $1.59$\\
	3.0 & $7.45\pm0.55$ & $1.72$ &  $5.1\pm0.11$ & $1.65$ & $4.63\pm0.09$ & $1.6$ & $4.94\pm0.12$ & $1.64$\\
	4.0 & $10.17\pm0.81$ & $1.82$ &  $6.82\pm0.22$ & $1.77$ & $6.06\pm0.16$ & $1.77$ & $5.75\pm0.14$ & $1.72$\\
	5.0 & - & - &  $9.22\pm0.72$ & $2.01$ & $8.4\pm0.28$ & $1.87$ & $6.93\pm0.18$ & $1.84$\\
	\hline
	\hline
	\multicolumn{9}{|c|}{fit for $L_{1}$: $r_{0}=a+b ~(1+z)+c~(1+z)^2$, with $a, b, c =[5.84, -1.47, 0.46]$}\\
	\hline
  \end{tabular}
  \caption{Same as the previous table, this time for the AGN obtained with Mod
    II (lower panel of Figure \ref{fig:corr_length_halo_ModI_ModIII}).}
  \label{table:r0_gamma2}
 \end{center}
\end{table*}

In this subsection we compare the AGN clustering with the clustering of the
Millennium haloes.  In our model, BHs are allowed to accrete cold gas only
during merger events, which are experienced mainly by the galaxies sitting at
the centers of FOF haloes. As discussed above, only a small fraction of AGN
can be hosted by satellite haloes. Due to
this uncertainty in the quasar pair regime, we focus in the present work on
the clustering on intermediate and large scales, and we refrain from drawing
strong conclusions from the results at scales much smaller than the average
halo separation.

In Figure~\ref{fig:corr_func_halo_ModI}, we show at different redshifts the
two-point correlation function of the AGN population, divided in four
luminosity bins depending on their intrinsic bolometric luminosity. This is
compared with the two-point correlation of the FOF haloes, divided into two
bins according to their virial mass. The AGN shown in this figure have been
obtained using Mod II for the lightcurve.  The corresponding correlation
lengths are shown in the lower panel of
Figure~\ref{fig:corr_length_halo_ModI_ModIII}. In the upper panel of the same
figure the correlation lengths of the AGN obtained using Mod I are plotted,
also divided in four luminosity bins. In the analysis of the results, we allow
the exponent $\gamma$ of the power-law ansatz for the correlation function to
vary in each fit. The values of $r_{0}$ and $\gamma$ for the two models thus
obtained are given in Tables \ref{table:r0_gamma1} and
\ref{table:r0_gamma2}. We also fitted the brightest bin with a quadratic
function ($r_{0}(z)=a+b ~(1+z)+c~(1+z)^2$) to compactly summarize the results,
and the values of the coefficients are given at the end of each table.

\begin{figure}
\begin{center}
	\includegraphics[width=0.45\textwidth]{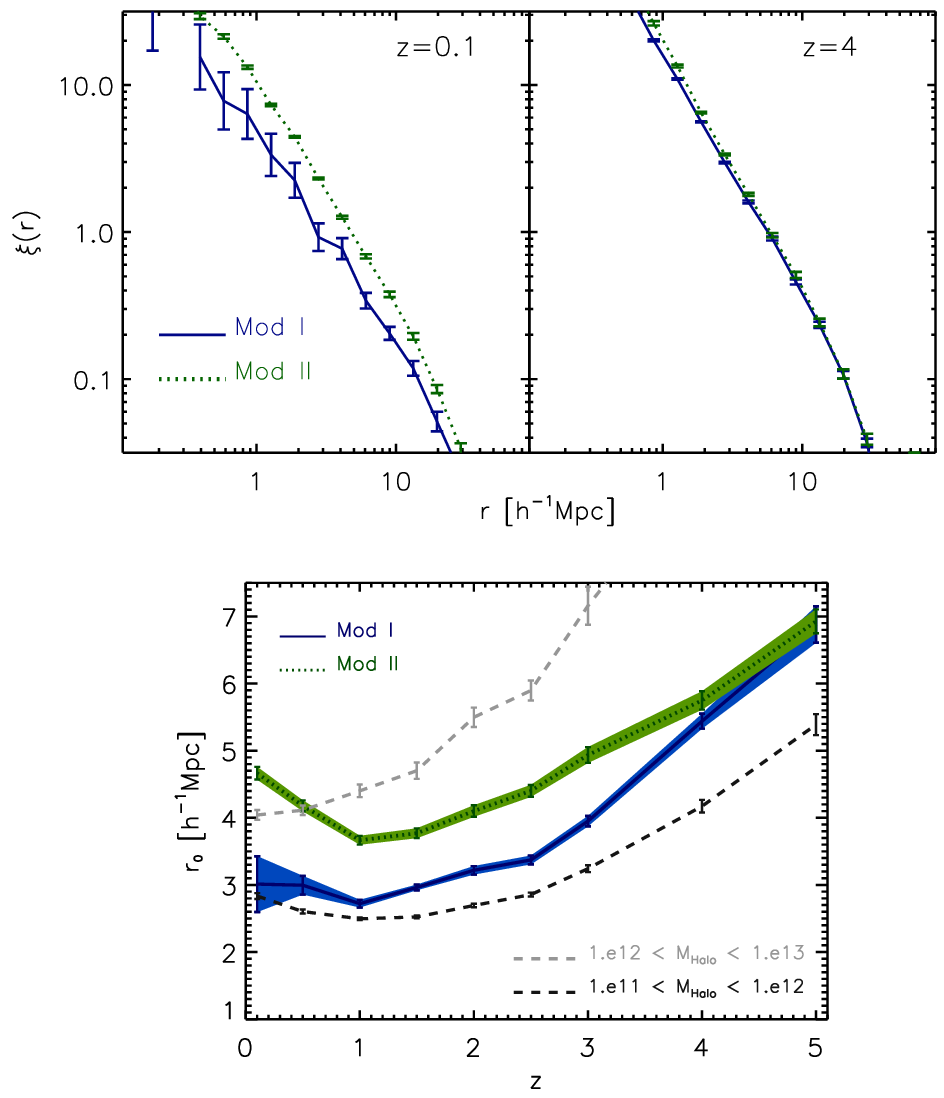} 
        \caption{We compare here the correlation function of faint AGN
          ($L_{\rm Bol}<10^{10} \rm{L}_{\odot}$) obtained using Mod I (solid
          blue line) and Mod II (dotted green line). We show the result at
          very high redshift, where there is no difference in the two models,
          and at low redshift, were the difference becomes significant. In the
          lower panel the corresponding correlation function is shown as a
          function of redshift, and the correlation of FOF haloes is shown for
          reference.}
        \label{fig:corr_faintAGN_halo_ModI_ModIII}
\end{center}
\end{figure}

Comparing the values of the correlation lengths obtained with the two models,
we do not find significant differences, except for the faintest AGN ($L_{\rm
  Bol}<10^{10}\rm{L}_{\odot}$). An enlarged view of the behavior of the
correlation strength of these faint objects obtained with Mod I (solid blue
curve) and with Mod II (dotted green curve) is shown in Figure
\ref{fig:corr_faintAGN_halo_ModI_ModIII}.  While at high redshifts there is
hardly any difference between the two models, at low redshift the faint
objects obtained with Mod II are much more strongly clustered. This is because
most of the population is composed of large BHs that are accreting at low
$\fedd$ (as shown in Figure \ref{fig:Edd_fraction}) and that are hosted by
large haloes.  In the lower panel of Figure
\ref{fig:corr_faintAGN_halo_ModI_ModIII}, we see that the correlation length
of the faint objects obtained with Mod II is comparable to the ones of haloes
with $M_{\rm{Vir}} \approx 10^{12}- 10^{13} \rm{M}_{\odot}$, while faint
objects obtained with a pure Eddington-limited accretion model are sitting in
haloes of much lower mass.  Observational clustering measurements have been
used in recent years to estimate the typical halo masses that host quasars
\citep[e.g.,][]{porciani04, grazian04, croom05}. This is usually done by comparing the bias
of observed quasars with the halo bias obtained from analytical estimates
\citep[e.g.,]{mo96, sheth99}.  In the present work, the host halo mass is an
output of the simulation, and therefore we can directly examine the relation
between black hole mass, quasar luminosity and halo mass. In section
\ref{section:QSO_HALO}, we exploit this for a direct study of the dark
environment of luminous BHs.

Based on Figure \ref{fig:corr_length_halo_ModI_ModIII}, it seems that the
redshift-evolution of the clustering of quasars is consistent with the
redshift-evolution of the clustering of dark matter haloes (quasars of a given
luminosity reside at all times in haloes of a fixed mass). Again, the only
substantial difference to
this trend is for the faint objects obtained with Mod II: since their
clustering is more constant with redshift, it implies that their typical host
halo mass changes with redshift.

\subsection{Luminosity dependence of AGN clustering and comparison with
observational data}  \label{sec:comp_obs}

In this subsection we first examine the dependence of AGN clustering on luminosity,
 looking at the global population, and then considering a subsample that
would be observable in the optical band.

Observationally, quasar clustering seems not to depend significantly on
luminosity \citep[e.g.,]{porciani04, croom05, daAngela08}. Only
\citet{shen08b} found indications of a luminosity-dependence of the clustering
when they compared the two-point correlation of their $10\%$ brightest objects
with the rest of the sample. Figure~\ref{fig:corr_length_halo_ModI_ModIII}
 provides information on how the correlation length evolves with
luminosity in our models. Except for the faintest bin (see Figure
~\ref{fig:corr_faintAGN_halo_ModI_ModIII}), there is not a substantial
difference between the two models, as pointed out before.  In both models we
see some moderate evolution with luminosity, and in particular, in both cases
the brightest quasar bin is substantially more strongly clustered than the
lower luminosities. 

Note that in this analysis a very large range in luminosities is covered
($\approx 5\,{\rm dex}$ in luminosity, corresponding to $\approx 12.5$
absolute magnitudes). Observationally, the accessible luminosity range is
always much smaller than that.  To give predictions that can be compared with
future observations, we now extract from the global AGN population sub-samples
of optically visible bright AGN.  First of all, to account for obscuration, we
calculate the fraction of objects that would be visible in the optical using
the `observable fraction' from \citet{hopkins07a}. This gives, as a function
of luminosity, the probability for an object to be seen in a given band:
\begin{equation}\label{eqn:obs_fraction}
f(L)=f_{46} \left ( \frac{L}{10^{46}\rm{erg s}^{-1}} \right ) ^{\beta},
\end{equation}
where $f_{46}=0.260$ and $ \beta = 0.082)$ for the B-band.

To convert from bolometric luminosity to B-band luminosity, we used the
bolometric corrections again from \citet{hopkins07a}:
\begin{equation} \label{eqn:bolometric_correction} \frac{L_{\rm bol}}{L_{\rm
      band}}= c_{1}\left( \frac {L_{\rm bol}}{10^{10}L_{\odot}} \right)^{k_1}
  + c_{2}\left( \frac {L_{\rm bol}}{10^{10}L_{\odot}} \right)^{k_1},
\end{equation}
where $(c1,k1,c2,k2)$ are respectively $(6.25,-0.37,9.0,-0.012)$ for the
B-band.

\begin{figure}
        \includegraphics[width=0.48\textwidth]{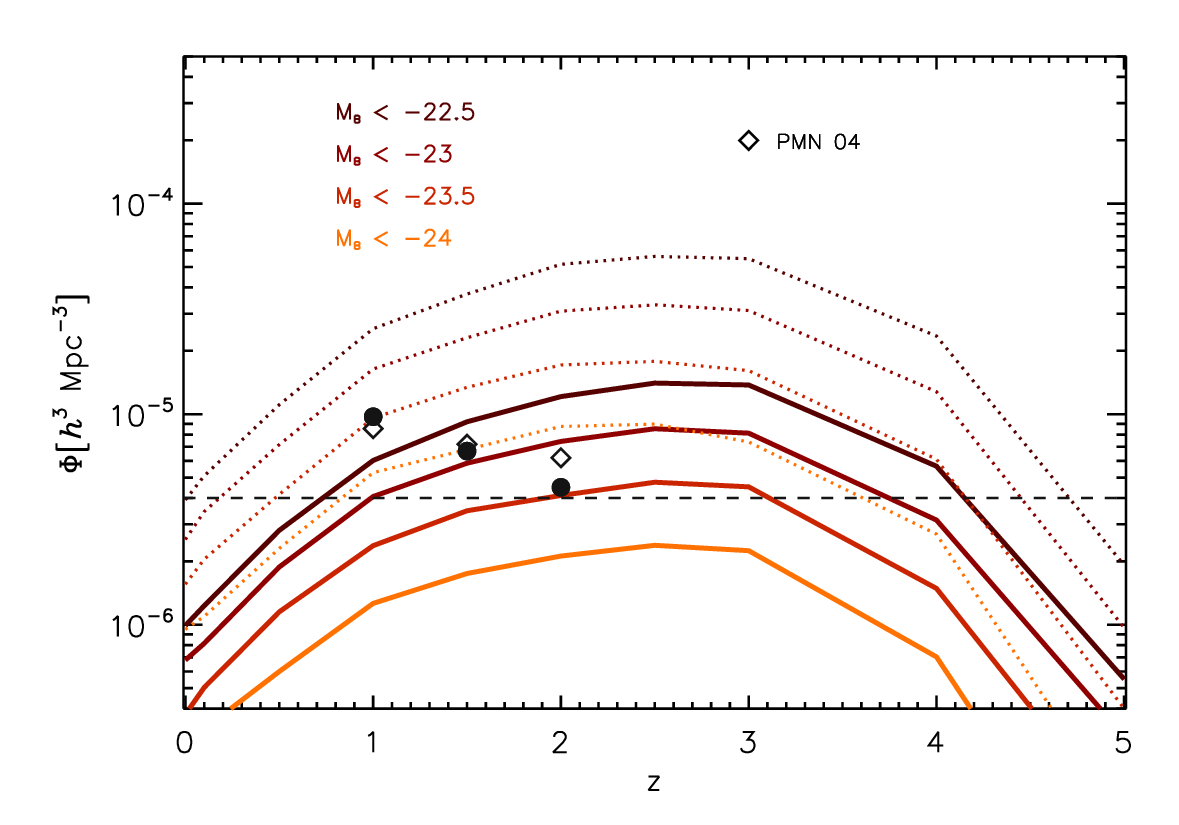}
        \caption{Space density as a function of redshift for four subsamples
          selected with B-band magnitude cuts as indicated on the plot. The
	solid lines are give the space density when the possible obscuration is
taken
	into account. If we allow all our objects to be optically visible, we obtain the
	space densities described by the dotted curves. The
          dashed line marks the point below which we have less than $500$
          objects remaining in in the Millennium simulation volume. The open
          diamonds are the observed values from \citet{porciani04}, obtained
          in different magnitude ranges depending on the redshift (see text
          for details). The number densities obtained with our model using the
          same magnitude ranges and accounting for obscuration are indicated with the filled circles.}
        \label{fig:space_density_BBand}
\end{figure}

\begin{figure}
\begin{center}
        \includegraphics[width=0.42\textwidth]{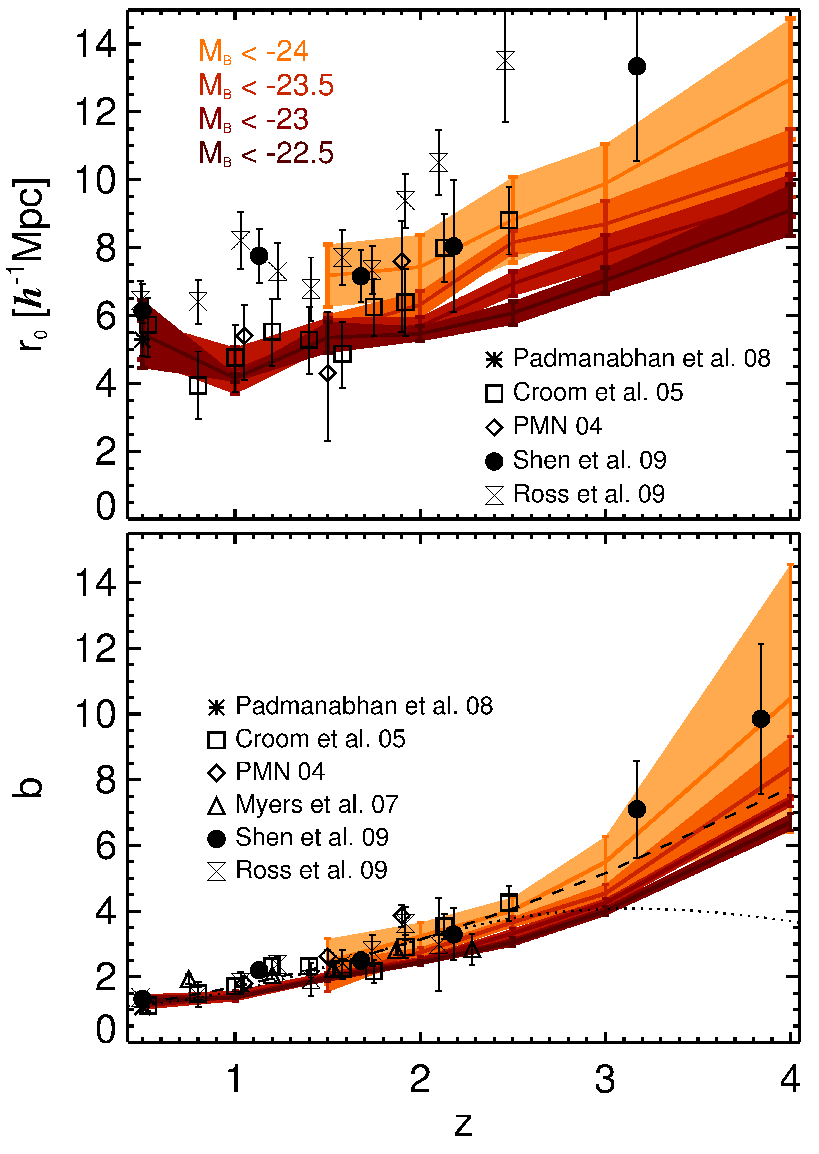}
        \caption{Correlation length (top panel) and bias (lower panel)
          for the AGN selected using the cuts of Figure
	\ref{fig:space_density_BBand} (neglecting the effects of obscuration). Due to
	lack of enough objects, the clustering properties of the two brightest bins are
	calculated only down to $z=1.5$. Our
	predictions are plotted together with observational data (for the
	\citet{shen08a}, we included their lower estimates).  For the bias, the
          dotted line is the prediction of \citet{hopkins07b} and the
          short-dashed line is the best fit from \citet{croom05}. }
        \label{fig:corr_length_bias_BBand}
\end{center}
\end{figure}

\begin{figure}
\begin{center}
        \includegraphics[width=0.45\textwidth]{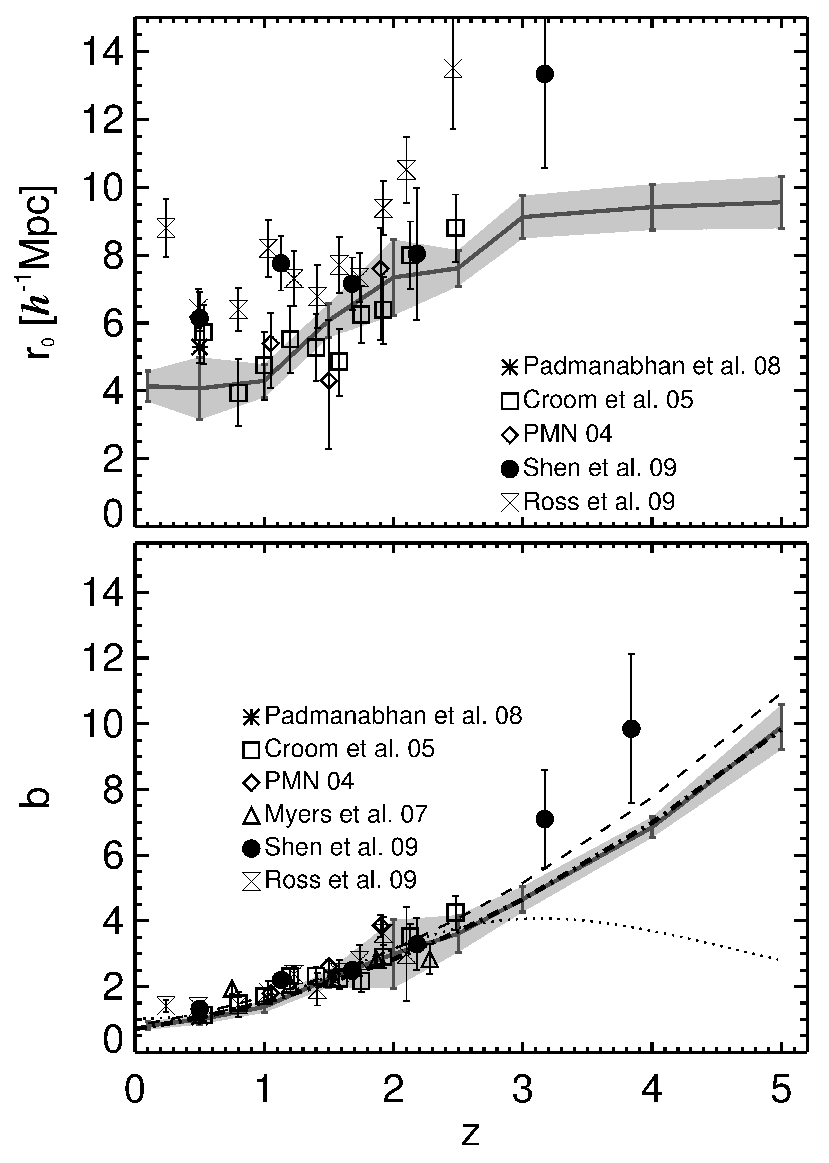}
        \caption{Correlation length (top panel) and bias (lower panel) for
          $L_{\ast}$ quasars. The gray line is our prediction (with errors
	enclosed in the grey area). The observational data are the same of Figure \ref{fig:corr_length_bias_BBand}. 
	A fit to our
          predicted bias as a function of redshift is given in equation
     	  (\ref{eqn:best_fit_bias_Lstar}). }
        \label{fig:corr_length_bias_Lstar}
\end{center}
\end{figure}

In Figure \ref{fig:space_density_BBand} we show as a function of redshift the
number density of our simulated AGN for different luminosity cuts (solid lines).
 In order to directly compare our
number densities with the values inferred from observational data used for
clustering measurements, we calculated in the same figure the number density
of objects in the magnitude ranges given by \citet{porciani04} at three
different redshifts: the values of $M_{\rm min}$ and $M_{\rm max}$ are
$[-25.32,-21.72]$ at $z \sim 1.0$, $[-25.97,-22.80]$ at $z \sim 1.5$, and
finally $[-26.44,-23.37]$ at $z \sim 2.0$ (see their Table 1). Note that their
value are in $b_{J}$, and to convert from $B$ to the $b_{J}$-band we used the
relation given by these authors in their Appendix 1, where
$M_{B}=M_{b_{J}}+0.07$. In the Figure, our points are the black dots, while
the numbers quoted by \citet{porciani04} are shown with diamonds (the errors
quoted by these authors are comparable to the size of the symbol, and
therefore are omitted). The agreement is quite good, even though we slightly
underestimate the number of bright quasars at $z=2$, as expected (see the
bright-end of the luminosity function at this redshift in Figure
\ref{fig:Lum_func_ModI_ModII} ).
 In Figure \ref{fig:space_density_BBand} we also show the number density of our simulated AGN for the
same luminosity cuts, but without accounting for obscuration
(dotted lines). As described above, we account for obscuration by calculating
for each object its probability of being optically visible and then by randomly
extracting objects that satisfy the imposed condition. Since this probability is
a weak function of luminosity, and since clustering analysis is
independent of random sampling, for our study we ignore the effect of
obscuration. This allows us to push the analysis to brighter magnitude cuts,
since for a statistically-accurate clustering analysis we need at least few hundred objects 
(the dashed horizontal line shows the point at which, in the full simulation volume, we
cannot expect more than $500$ objects). 

The correlation lengths of the AGN selected with these luminosity
cuts  are shown in Figure \ref{fig:corr_length_bias_BBand}. 
We see that at low and
intermediate redshifts the correlation length and the bias  depend  weakly 
on luminosity when a narrow range of luminosities is
examined. Since bright quasars are always powered by BHs accreting close to
the Eddington limit, it seems difficult to use quasar clustering observations
to disentangle between different light-curve models, unless much larger luminosity
ranges are probed. The present observations indicate however that, over the
range of luminosities observed, quasars reside in haloes of similar
masses. Based on our results, we conclude that the lack of a significant
dependence of clustering on luminosity is not primarily a result of invoking
 lightcurve models with a wide distribution of Eddington ratios, but rather
arises because in a merger-driven scenario
there is a small scatter in the typical halo mass hosting quasars close to
their peak luminosity.

In Figure \ref{fig:corr_length_bias_BBand} we
added observational data from several works, to qualitatively compare our
results with observations. We stress though that the error bars in these figures
are calculated to describe the effect of cosmic variance
as described in Section \ref{sec:intro_clustering}; since we are here ignoring the effect of obscuration, thus
improving our statistic, a direct comparison with the error bars given by
observational works is not possible.

   Most of the observed quasars have a typical
magnitude around $M^{*}_{b_{J}}$ \citep{croom05}, with faint limits that
strongly depend on redshift (at very low redshifts surveys can
reach fainter magnitudes, whereas at very high redshifts the limiting
magnitudes can be higher than $M_{\ast}$). At $z\la 1$ the faintest observed magnitudes are
$M_{B} \approx -22$, going up to $\approx -24$ at $z\sim 2-3$. Since each
observational study uses different magnitude cuts, we can not do a detailed
comparison with all the observations available, but overall our results
for the values of the correlation length and the bias and their evolution with
redshift are in good agreement with the observational results. 

We also compared observational data with simulated quasars around  $L_{\ast}$,
calculated using equation 9 from
\citet{hopkins07a}, and selecting objects with an
intrinsic luminosity larger than $L_{\ast}/0.5\, {\rm dex}$ (which corresponds
to a minimum luminosity approximately $1.2\,{\rm mag}$ fainter than
$M_{\ast}$). Our predictions for the correlation length and the bias for
$L_{\ast}$ objects as a function of redshifts are shown in
Figure~\ref{fig:corr_length_bias_Lstar}, again together with the available
observational data.  The discrepancies
with \citet{shen08b} for the correlation length can be due in differences in the
calculation of this quantity (as already mentioned, here we do not fix the value
of $\gamma$). For the bias,
we show also the best fit from \citet{croom05} and the prediction of
\citet{hopkins07b}. The latter was probably fitted only up to $z=3$, thus
explaining the turn-over at redshifts above $3$ that seems to not be
consistent with the trend shown by the observations. A good approximation to
our prediction for the bias is given by the fitting function
\begin{equation}\label{eqn:best_fit_bias_Lstar}
b(z)=0.42+0.04(1+z)+0.25(1+z)^2.
\end{equation}

Quasars with luminosities around $L_{\ast}$ are typically objects very close
to their peak luminosity, therefore correspond to objects accreting at high
Eddington ratios.  As mentioned before, we cannot use these results as a sensitive test
of our lightcurve models. However, the good agreement with observations
indicates that our merger-triggered BH accretion model predicts a spatial
distribution of quasars that is consistent with observations. 
This is a prediction of a consistent model of the joint evolution of dark
matter, galaxies and black holes, evolving $\Lambda$CDM initial conditions
from high redshift to the present. While the parameters of the
semi-analytic model had been tuned to fit the bulk $z=0$ properties of the BH
population and the AGN luminosity function as a function of redshift,
information on clustering had not been considered in the construction of the
model, and therefore must be regarded as genuine model predictions.

\section{BHs, Quasars and their dark environment}\label{section:QSO_HALO}

\begin{figure*}
\begin{center}
        \includegraphics[width=0.98\textwidth]{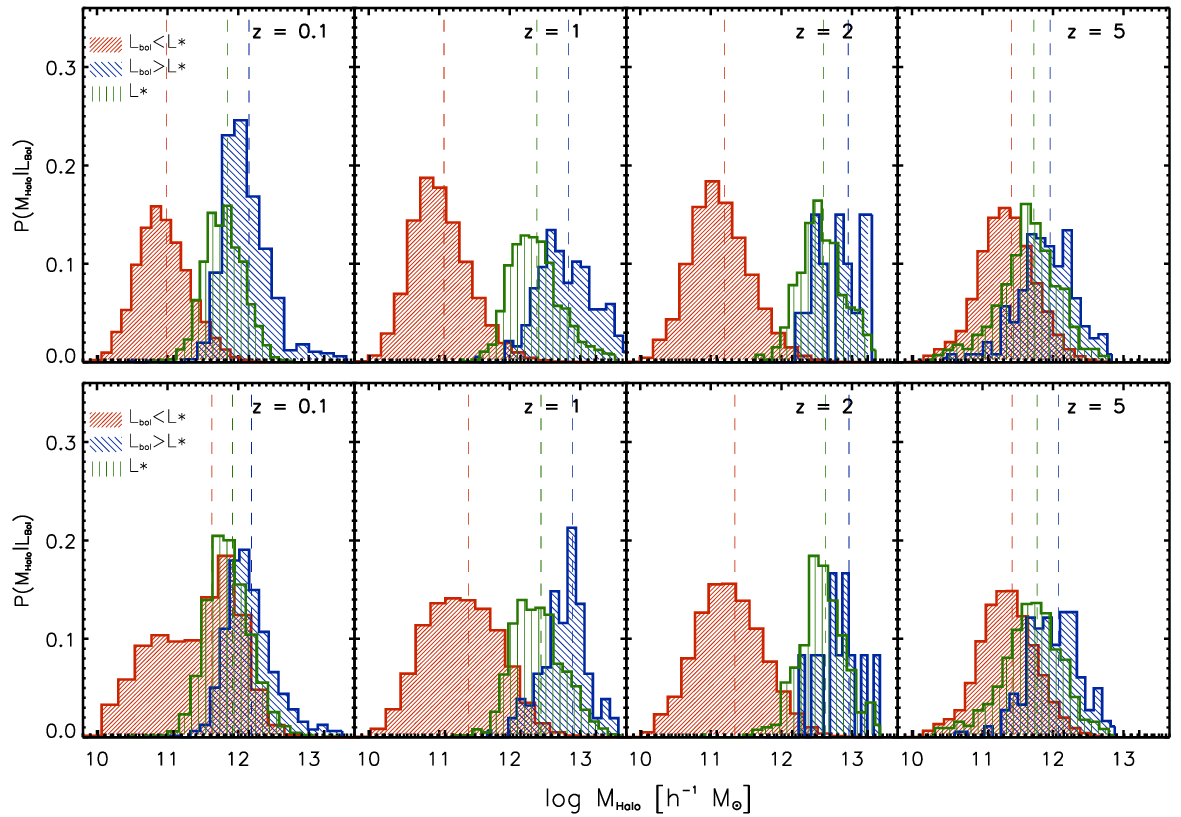}
        \caption{Distribution of dark matter halo masses hosting faint-AGN,
          bright AGN and $L_{\ast}$ quasars. The vertical dashed line
          indicates the median of the distribution for each luminosity bin, and we refer the reader to
          the legend on the plot for details in the color/pattern-coding. The
          AGN have been obtained using Mod I (upper panel) of Mod II (lower
          panel) for the lightcurve, respectively.}
        \label{fig:Halo_dist}
\end{center}
\end{figure*}

In this section we explore directly the connection between BHs, quasars and
their dark matter environment. As in our simulations the dark matter halo
merger trees are the backbone upon which the baryonic component is treated, we
can also use them to study the dark environment of our AGN. This in particular
allows tests of the validity of the approach typically adopted in the
interpretation of observational quasar clustering results
\citep[e.g.][]{porciani04, croom05}, where the observed quasar bias is
compared with the halo bias predicted by analytical halo models
\citep[e.g.][]{mo96, sheth99}.

The mass distribution of the haloes hosting AGN of given luminosities,
$P(M_{\rm Halo}|L_{\rm AGN})$, is shown in Figure \ref{fig:Halo_dist}.  The AGN
are here sub-divided into a faint and a bright sub-sample, depending on their
bolometric intrinsic luminosity. The cut in bolometric luminosity is here
$L_{\ast}$, calculated in the same way as for section \ref{sec:comp_obs}.
Based on the results on the Eddington ratio distribution (see Figure
\ref{fig:Edd_fraction}) and on the clustering, we expect the distribution of
the masses of the haloes hosting bright AGN to be similar both for Mod I and
Mod II. The main difference should be in the distribution of haloes hosting
faint AGN: in the Eddington-limited model, the faint AGN population is
composed of small-mass BHs accreting at Eddington, whereas in the model that
includes a long quiescent phase the faint-AGN population at low redshifts
includes also quite massive BHs accreting at low Eddington ratios.  

In Figure~\ref{fig:Halo_dist} we indeed see that for Mod I there is a direct
proportionality between the luminosity of the AGN and the mass of the host
halo: the brighter the AGN, the larger the BH and the host halo. Instead, for
Mod II most of the low-luminosity AGN at low redshifts are hosted by more
massive haloes, i.e., massive BHs accreting at low Eddington ratio.  In the
same figures we also plot the mass distribution of haloes hosting $L_{\ast}$
quasars. To get a large enough sample, at any given redshift we included
objects in a range of $\pm 0.5\,{\rm dex}$ around $L_{\ast}$. The similar
behavior of haloes hosting $L_{\ast}$ quasars in both models suggests
that $L_{\ast}$ objects are mainly BHs accreting close to the Eddington limit.

\begin{figure}
        \includegraphics[width=0.45\textwidth]{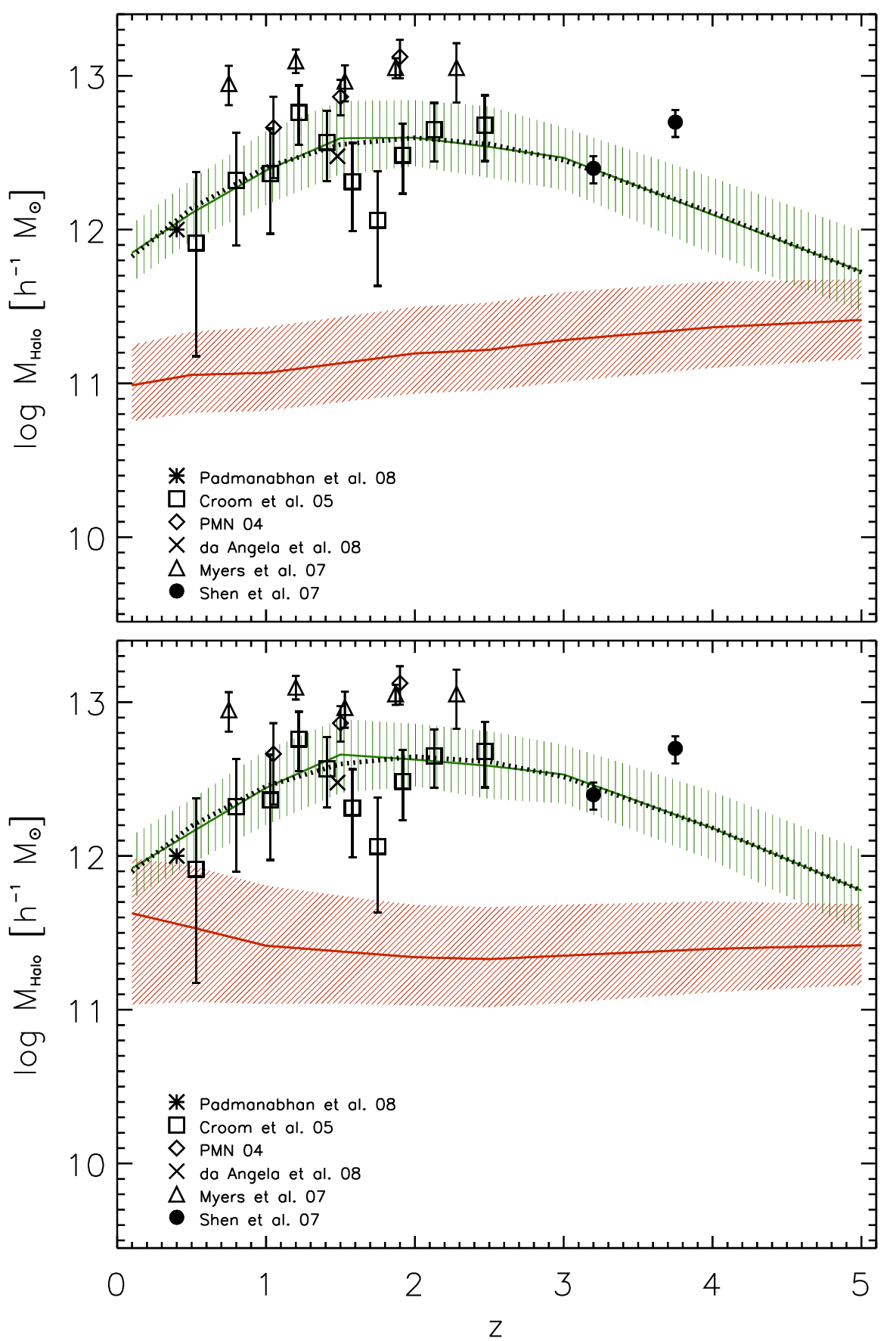}
        \caption{In these two panels we show the redshift evolution of the
          median mass of dark matter haloes hosting AGN of different
          luminosities from the previous Figure (\ref{fig:Halo_dist}). For
          clarity in the plot, we only show the values obtained for objects
          with $L_{\rm{Bol}} < L_{\ast}$ and with $L_{\rm{Bol}} \sim
          L_{\ast}$. The dotted black curve shows the best fit to the evolution
of the typical host mass of  $L_{\ast}$ quasars.
	The contours indicate the $25$ and $75$ percentiles.  We
          overplot here estimates obtained by different groups who examined
          the clustering properties of observed quasars (see legend on the
          plots).}
	\label{fig:median_host_haloes}
\end{figure}

The mean values of the distributions are shown as a function of redshift in
Figure \ref{fig:median_host_haloes}. In recent years many groups have analyzed
the clustering properties of quasars to estimate the typical mass of their
host haloes, at low- \citep{padmanabhan08}, intermediate- \citep{croom05,
  porciani04, daAngela08, myers07} and high- \citep{shen07} redshifts.  These
works used quasars observed with SDSS and $2$dF, with a typical luminosity
around $L_{\ast}$ (except for the very high-redshifts measurements).  The
masses of the dark matter haloes hosting quasars estimated by these groups are
overplotted in Figure~\ref{fig:median_host_haloes}. Almost all these estimates are in
the range predicted by our model: the typical halo mass hosting $L_{\ast}$
quasars seems to grow up to $z \approx 1.5-2$, and then it decreases again at
higher redshifts.  To compactly represent our
simulation results, we fitted our results with a cubic function
\begin{equation}\label{eqn:cubicfitMHalo}
  M_{\rm halo} = a_{0}+ a_{1} z+ a_{2}z^{2} + a_{3}z^{3}, 
\end{equation}
with $a_{i} = [ 11.873;0.944;-0.318;0.026]$ for the second panel of Figure
\ref{fig:median_host_haloes} (the values of these coefficients are similar for
the fit of the $L_{\ast}$ curve of the upper panel, which we omit for
brevity).

\begin{figure}
\begin{center}
	\includegraphics[width=0.48\textwidth]{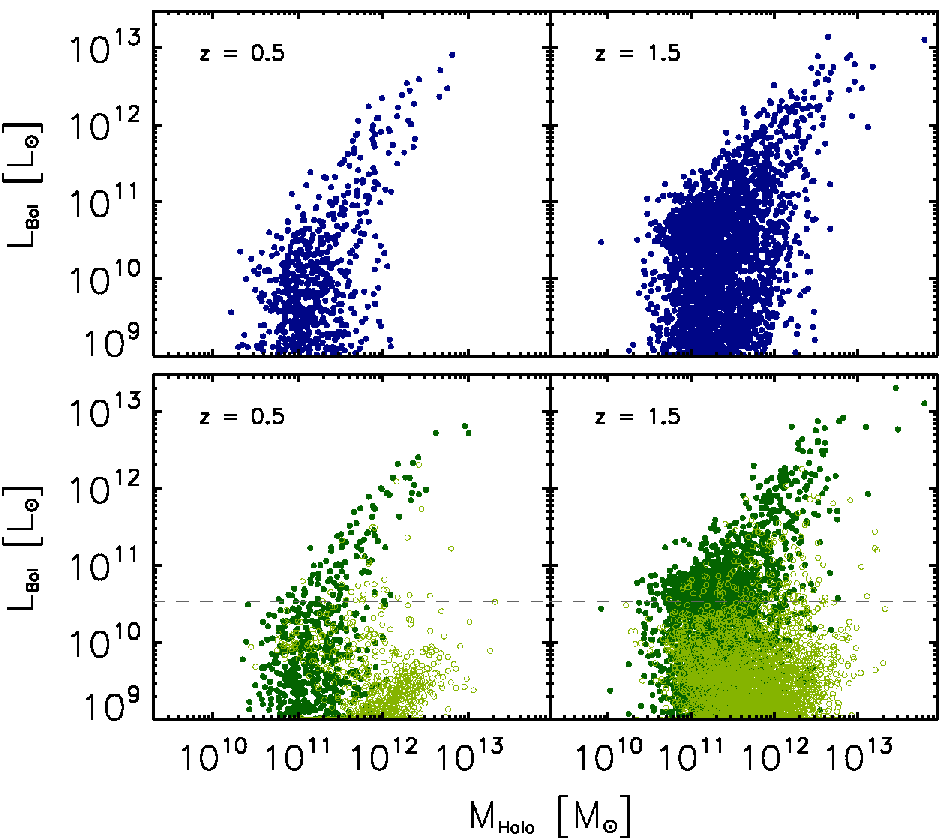}
	\caption{Relation of $L_{\rm{Bol}}$ of the AGN versus dark matter halo
          mass. In the upper panel, BHs accrete according to the Mod I
          lightcurve, while in the lower panel the predictions are produced
          using Mod II. While all very bright objects are BHs accreting close
          to the Eddington limit, the main difference between the two models
          lies in the faint objects, where we have a dense
          population of faint AGN hosted by large haloes (the light-green open circles in
	  the lower panel refer to AGN in the quiescent phase). For reference, the
          dashed line marks the Eddington luminosity corresponding to a BH
          mass of $10^{6}\, \rm{M}_{\odot}$}
	\label{fig:Lum_Halo_rel}
\end{center}
\end{figure}

Our results for bright quasars (objects around $L_{\ast}$) are also consistent
with the estimates of \citet{lidz06} and \citet{hopkins07b}, who calculate
that the typical mass of haloes hosting quasars is $\approx 4 \times 10^{12}
h^{-1} \rm{M}_{\odot}$. These authors argue that bright and faint quasars are
the same type of objects but seen in different evolutionary states, and
therefore their typical host halo mass should be similar. Since only the
brightest quasars are objects accreting at high $\fedd$, only for these
objects we expect a tight relation between the instantaneous luminosity and
the host halo mass. The relation between AGN luminosity and halo mass is shown
in Figure \ref{fig:Lum_Halo_rel}. Indeed, only for the very bright quasars
there is a direct proportionality between luminosity and halo mass. These are
objects that are close to their peak luminosity, have accreted most of the gas
available, and at this point their BH is tightly correlated with the mass of
the host halo (see also next figure). During the rising phase of the
lightcurve (even if BHs are accreting at Eddington), BHs are not yet strongly
correlated with the host halo, reflected in a lack of correlation between
quasar luminosity and halo mass.  During the decaying phase, Mod II produces a
dense population of faint objects sitting in massive haloes (see open circles
in Figure \ref{fig:Lum_Halo_rel}). 

\citet{white07} claimed that the very high bias observed for high-redshift
quasars implies a small dispersion in the above relation. Estimates of high
Eddington ratios for bright objects at high redshifts \citep{kollmeier06,
  shen08a} indeed seem to support that for very bright objects a tight
relation exists between quasar luminosity and halo mass
\citep{fine06}. However, we would like to point out that just looking at the
bright quasar population it is not sufficient to distinguish between different
lightcurve models.

\begin{figure*}
\begin{center}
	\includegraphics[width=1.0\textwidth]{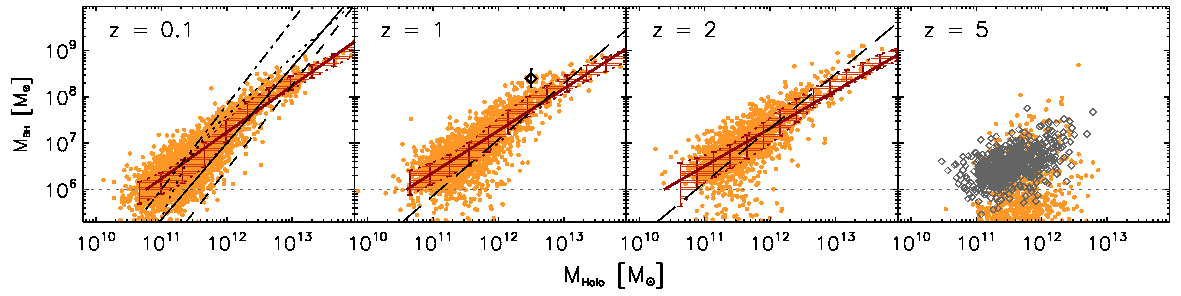}
	\caption{$\MBH - M_{\rm{Halo}}$ relation for BHs sitting in central
          galaxies.  The points are our simulated objects, and the red line is
          the best-fit assuming a linear relation. The filled region encloses
          the $25$ and $75$ percentiles. For reference, we show at $z=0.1$ the
          result that \citet{ferrarese02} obtained at $z=0$ assuming
          $v_{\rm{vir}} = v_{c}$ (dashed line), $v_{c} =1.8 v_{\rm{vir}}$
          (dot-dashed line) and the prescription from \citet{bullock01} for
          this relation (solid line). At $z=0.1$ we show also the result from
          \citet{shankar06} (dotted curve). The point at $z=1$ is the
          zero-point of this relation obtained by \citet{fine06}. The dashed
          lines at $z=1$ and at $z=2$ are from \citet{colberg08} (for $z=2$ we used their result at $z=3$). 
	  The horizontal dashed line marks $\MBH = 10^{6}\, \rm{M}_{\odot}$, which
          is approximately our resolution. This plot was obtained assuming Mod
          I for the lightcurve, but the result does not change using Mod II,
          since the final BH masses are the same. The diamonds at $z=5$ show
          the relation between BH mass and halo mass if BHs accreted the
          available mass instantaneously.}
	\label{fig:BH_Halo_rel}
\end{center}
\end{figure*}

The observed scaling relations between BH masses and different properties of
the host galaxy have suggested the possibility of a more fundamental
connection between the mass of the BH and the host system. Using measurements
of stellar velocity dispersions and assuming a relation between this quantity
and the circular velocity of the galaxy and the BH mass, \citet{ferrarese02},
\citet{baes03} and \citet{shankar06} estimated how the BH mass could be
connected to the dark halo mass in the local universe.  At higher redshifts
these estimates are of course more problematic, because studies of the stellar
kinematics are unavailable and we also are not certain yet how the $\MBH
-\sigma$ relation evolves with redshift. \citet{fine06} explored the relation
between BHs and quasar host haloes at $z=0.5-2.5$ using BH virial masses
estimates from the width of broad emission lines and DM halo mass obtained
from quasar clustering from \citet{croom05}. In Figure~\ref{fig:BH_Halo_rel},
we plot the $\MBH - M_{\rm{Halo}}$ relation for our simulated BHs. We include
here only BHs residing in central galaxies of FOF haloes.  This is because in
our model only central galaxies can merge, and therefore it is mainly BHs
 hosted by FOF haloes that can grow (the results of \citet{li06} indicate that this could be supported by
observations) . Indeed, we find a well-defined relation which gets tighter with
decreasing redshift. In Paper I we already showed this relation at redshift
$z=0$ and we found good agreement with other works \citep{ferrarese02, baes03,
  shankar06}. Here we overplot the results of \citet{ferrarese02} and
\citet{shankar06} at $z=0.1$ for reference; at $z=1$ we overplot the
zero-point in the relation estimated by \citet{fine06} ($\MBH=10^{8.4\pm0.2}
\rm{M}_{\odot}$ for a halo of $M_{\rm{halo}}=10^{12.5} \rm{M}_{\odot}$) and at
$z=1$ and $z=2$ the results from direct hydrodynamical simulations of
\citet{colberg08} (for $z=2$ we used their result at $z=3$).

Note that the fact that BHs need to accrete most of the available gas before
they `sit' on the above relation could be influenced at high redshifts by the
resolution limit of the Millennium simulation, which does not resolve low-mass
haloes below $\sim 10^{10}\,h^{-1}{\rm M}_\odot$.  We will explore this
high-redshift behavior in more details in future work.

\subsection{Duty cycle}

The time BHs spend shining as quasars is still an open question \citep[see
review by][]{martini04}. The definition itself of a `quasar lifetime' is
somewhat ambiguous. Observationally it is defined as the time BHs spend
shining at luminosities higher than some limit (for quasars, the usual
definition is the time an active nuclei shines with a B-band magnitude $M_{B}
< -23 \, {\rm mag}$). Theoretically, it can be defined in a simpler way as the
total time a BH shines at high Eddington ratio. The quasar lifetime is often also
 simply defined through the duty cycle, which is given by ratio of the quasar number
density and the number density of the haloes that can host them: $t_{q} \sim
t_{Hubble} n_{q}/n_{Halo}$ \citep[e.g.,][]{adelberger05}. 

\citet{haiman01} and \citet{martini01} suggested to use the observed quasar
clustering to estimate the quasar lifetime, upon the assumption that a
monotonic relation exists between quasar luminosity and halo mass \citep[see
also][]{haehnelt98}.  \citet{adelberger05} pointed out that the theoretical
estimate of the duty cycle through clustering analysis depends on the Eddington
ratio distribution, on obscuration and on the scatter in the realtion between
quasar  luminosity and halo mass. As we have seen, the assumption of a tight
relation between luminosity and halo mass is overly simplistic for 
realistic lifetime models, and it is therefore interesting to use our simulations 
directly to examine the distribution of quasar lifetimes. 

\begin{figure*}  
\begin{center}    
	\includegraphics[width=0.70\textwidth]{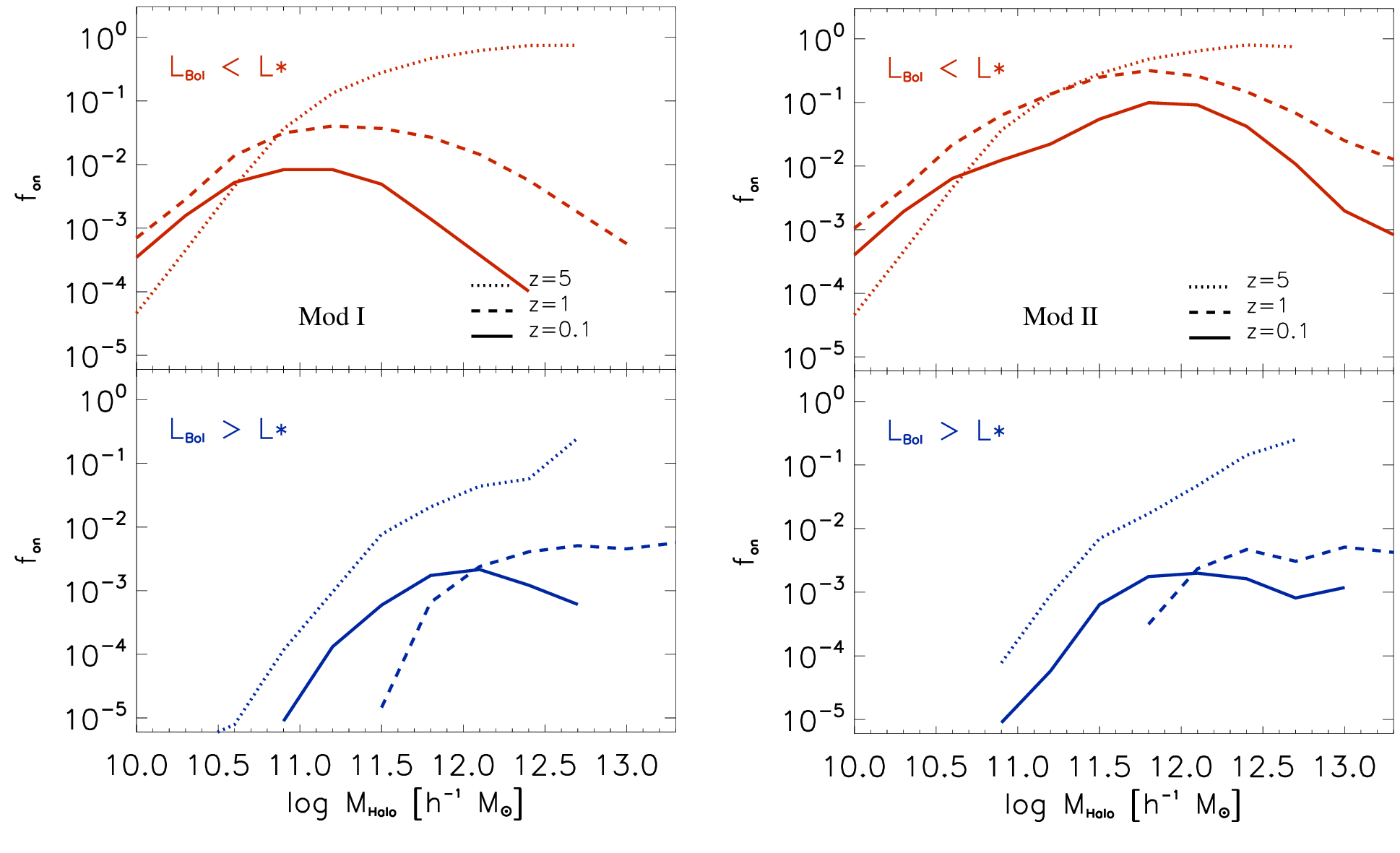}
        \caption{Fraction of active haloes (or duty cycle), as a function of
redshift, halo mass and AGN luminosity. We compare the results obtained for Mod
I (left panels) and Mod II (right panels).}
	\label{fig:duty_cycle}
\end{center}
\end{figure*}

In Figure~\ref{fig:duty_cycle} we show the fraction of active haloes (or duty
cycle), as a function of quasar luminosity, redshift and halo mass, for both Mod
I (left panels) and Mod II (right panels). At high redshifts massive haloes
have a very high duty cycle, i.e., most of haloes host a bright quasar. As
expected, the duty cycle evolves more strongly with redshift for the more
luminous AGN: by redshift $z=0.1$ only $\approx 0.1\%$ of the more massive
haloes host a quasar, and this result is independent on the lightcurve model
assumed. Again, the difference in the two models is in the faint AGN population:
the duty cycle of faint objects evolves strongly with redshift and mass for Mod
I, since at low redshift only the smallest haloes host an active BH. On the
other side, if the AGN lightcurve includes a long low-level phase, then at low
redshift also massive haloes are hosting a low-luminosity object.

Estimates of the quasar lifetime obtained from quasar clustering suggest
timescales of the order of $10^{7} - 10^{8} \rm{yr}$, depending on the
redshift.  At high redshifts ($z \geq 3.5$), \citet{shen07} estimated
lifetimes of the order of $30 \sim 600\, \rm{Myr}$, while at $2.9 \leq z <
3.5$ the estimated range decreases to $4 \sim 50\,
\rm{Myr}$. \citet{porciani04} suggest $t_{q} \sim 10^{7} \rm{yr}$ at $z \sim
1$, and values approaching $10^{8} \rm{yr}$ at higher redshifts. As we
approach low redshifts and the local universe, the quasar lifetimes seem to
decrease: \citet{padmanabhan08} suggest values $< 10^{7} \rm{yr}$ for their
sample of quasars at $0.2 < z < 0.6$. As we have shown in Figure
\ref{fig:duty_cycle}, a strong evolution of the quasar lifetime is also expected
from our models: at intermediate-high redshifts our results are compatible with
lifetimes of a few $10^8 \rm{yr}$, but the detailed evolution of the duty cycle
also depends stongly on the range of host halo mass considered.

\section{Conclusions}

In this series of papers we investigate semi-analytic models for BH accretion
and quasar emission in the context of a comprehensive galaxy formation model
developed for the Millennium Simulation.  The physical scenario for BH growth
we study is based on the model for BH accretion from \citet{kauffmann00}, as
revised by \citet{croton06a}, which assumes that galaxy mergers are the
primary physical mechanism responsible for efficiently feeding central BHs.
In Paper I \citep{marulli08} we used the most recent observations of the local
BH population to test basic predictions of the model for the local BH
demographics, testing also different theoretical models for the quasar
lifetime with goal to reproduce the observed quasar luminosity function. We
found an overall good agreement between the predicted and the observed BH
properties, and that the faint-end of the observed luminosity function can be
better reproduced when a quasar lightcurve model is adopted that includes long
quiescent accretion after an Eddington-limited accretion phase.

In the present work we used the spatial distribution of active BHs as a
further test of our model for BH accretion. Throughout the paper, we compared
the results obtained adopting two different theoretical models for the quasar
lifetime: pure Eddington-limited accretion (Mod I), and a model in which
Eddington-limited accretion is followed by a long, weak accretion phase (Mod II),
modeled after \citet{hopkins05}.  The main difference between the predictions
of the two models is in the faint-end of the luminosity function. The long
low-luminosity accretion phase allowed by Mod II gives rise to a large
population of massive BHs that at low redshifts are accreting at low Eddington
ratios, in agreement with the observational results of, for example,
\citet{heckman04} and \citet{netzer07}, who found that in the local universe only BHs
with mass $\la 10^{7} \rm{M}_{\odot}$ are experiencing high-efficient
accretion. As also recently pointed out by \citet{hopkins08}, it is only by
studying the properties of the faint AGN population that the quiescent phase
described by \citet{hopkins05} can be tested.

Independent of the model adopted for the lightcurve, the two-point correlation
function of our simulated AGN can be approximated by a single power-law in the
range $0.5 \la r \la 20~ h^{-1} \rm{Mpc}$.  The bias between AGN and the dark
matter is a strong function of redshift, but, at a given epoch, it is
approximately constant in the range $1.0 \la r \la 20 ~h^{-1} \rm{Mpc}$.  As
expected, the correlation lengths of AGN obtained with Mod I or Mod II differ
only for the faint population: the correlation length of faint AGN obtained
with Mod II is consistent with the correlation length of $10^{12}-10^{13}
h^{-1} \rm{M}_{\odot}$ haloes, whereas faint AGN obtained with Mod I exhibit
the same clustering as $10^{11}-10^{12} h^{-1} \rm{M}_{\odot}$ haloes.

Recent results from optical quasar surveys like SDSS and 2dFQSO have not found
evidence for a strong dependence of clustering on luminosity
\citep[e.g.,]{porciani04, croom05, myers07, daAngela08}, except for
\citet{shen08b} who detect an excess of clustering for their $10\%$ brightest
quasars.  Our results are consistent with these observations if we consider
only quasars with an intrinsic luminosity within the range probed by these
surveys. However, if we compare the clustering properties of AGN over a very
extended range of luminosity, then the correlation length becomes a moderately
strong function of luminosity and the value of the correlation length of the
faint population in particularly is seen to depend on the lightcurve model
assumed. The fact that the clustering of the observed quasars does not depend
on luminosity could be explained in two ways: quasars of different
luminosities are powered by BHs of the same mass that are in different stages
of their evolution, and/or the typical mass of haloes hosting quasars is
approximately constant for the luminosity range probed by observations. From
our results the second hypothesis seems to be clearly favoured. The mass range
of haloes hosting $L_{\ast}$ quasars is narrow enough that a significant
luminosity dependence of clustering cannot be detected with the current
observational samples, independent of the lightcurve model.

We also directly compared the clustering of our $L_{\ast}$ quasars with the
most recent observational data, and found very good agreement. Since quasars
at these luminosities are objects very close to their peak luminosity, and
therefore correspond to objects accreting at high Eddington ratios, we cannot, 
however, use these results as a sensitive test of our lightcurve
models. Nevertheless, the good agreement with observations indicates that our
merger-triggered BH accretion model predicts a spatial distribution of quasar
that is consistent with observations. This non-trivial outcome can be
viewed as a further success of the hierarchical galaxy formation paradigm, and
the cold dark matter hypothesis.

We note that a similar result for the luminosity dependence of AGN
clustering has been found in \citet{marulli09}, who analyzed  mock AGN 
Chandra catalogues constructed with the same semi-analytic model adopted 
in  this work. Furthermore, \citet{thacker08} have recently found very 
similar results modeling the AGN spatial properties in an 
hydrodynamical simulation.

In future work we will compare the merger-triggered quasar model with
alternative suggestions for the physical triggering mechanism of quasar
activity, such as disk-instabilities occuring in isolated galaxies. We expect
that quasar clustering statistics can here be a potentially powerful
discriminant to further constrain the viable physical models for the evolution
of supermassive black holes, and their co-evolution with galaxies.

\section*{Acknowledgments}

We thank Gabriella De Lucia and Andrea Merloni for many interesting discussions
and valuable suggestions. SB acknowledges the PhD fellowship of the
International Max Planck Research School in Astrophysics. LM, EB and FM acknowledge partial 
support by ASI contract I/016/07/0 
"COFIS", ASI-INAF I/023/05/0, ASI-INAF I/088/06/0.

\bibliographystyle{mn2e}
\bibliography{Bonoli_BH}

\begin{thebibliography}{}

\bibitem[\protect\citeauthoryear{{Adelberger} \& {Steidel}}{{Adelberger} \&
  {Steidel}}{2005}]{adelberger05}
{Adelberger} K.~L.,  {Steidel} C.~C.,  2005, \apj, 630, 50

\bibitem[\protect\citeauthoryear{{Alvarez}, {Wise} \& {Abel}}{{Alvarez}
  et~al.}{2008}]{alvarez08}
{Alvarez} M.~A.,  {Wise} J.~H.,    {Abel} T.,  2008, in {O'Shea} B.~W.,
  {Heger} A.,  eds, First Stars III Vol.~990 of American Institute of Physics
  Conference Series, {Black Hole Remnants of the First Stars}.
pp 432--434

\bibitem[\protect\citeauthoryear{{Baes}, {Buyle}, {Hau} \& {Dejonghe}}{{Baes}
  et~al.}{2003}]{baes03}
{Baes} M.,  {Buyle} P.,  {Hau} G.~K.~T.,    {Dejonghe} H.,  2003, \mnras, 341,
  L44

\bibitem[\protect\citeauthoryear{{Bajtlik}, {Duncan} \& {Ostriker}}{{Bajtlik}
  et~al.}{1988}]{bajtlik88}
{Bajtlik} S.,  {Duncan} R.~C.,    {Ostriker} J.~P.,  1988, \apj, 327, 570

\bibitem[\protect\citeauthoryear{{Barnes} \& {Hernquist}}{{Barnes} \&
  {Hernquist}}{1996}]{barnes96}
{Barnes} J.~E.,  {Hernquist} L.,  1996, \apj, 471, 115

\bibitem[\protect\citeauthoryear{{Best}, {Kauffmann}, {Heckman}, {Brinchmann},
  {Charlot}, {Ivezi{\'c}} \& {White}}{{Best} et~al.}{2005}]{best05}
{Best} P.~N.,  {Kauffmann} G.,  {Heckman} T.~M.,  {Brinchmann} J.,  {Charlot}
  S.,  {Ivezi{\'c}} {\v Z}.,    {White} S.~D.~M.,  2005, \mnras, 362, 25

\bibitem[\protect\citeauthoryear{{Bromm} \& {Loeb}}{{Bromm} \&
  {Loeb}}{2003}]{bromm03}
{Bromm} V.,  {Loeb} A.,  2003, \apj, 596, 34

\bibitem[\protect\citeauthoryear{{Bullock}, {Kolatt}, {Sigad}, {Somerville},
  {Kravtsov}, {Klypin}, {Primack} \& {Dekel}}{{Bullock}
  et~al.}{2001}]{bullock01}
{Bullock} J.~S.,  {Kolatt} T.~S.,  {Sigad} Y.,  {Somerville} R.~S.,  {Kravtsov}
  A.~V.,  {Klypin} A.~A.,  {Primack} J.~R.,    {Dekel} A.,  2001, \mnras, 321,
  559

\bibitem[\protect\citeauthoryear{{Carswell}, {Whelan}, {Smith}, {Boksenberg} \&
  {Tytler}}{{Carswell} et~al.}{1982}]{carswell82}
{Carswell} R.~F.,  {Whelan} J.~A.~J.,  {Smith} M.~G.,  {Boksenberg} A.,
  {Tytler} D.,  1982, \mnras, 198, 91

\bibitem[\protect\citeauthoryear{{Cattaneo} \& {Bernardi}}{{Cattaneo} \&
  {Bernardi}}{2003}]{cattaneo03}
{Cattaneo} A.,  {Bernardi} M.,  2003, \mnras, 344, 45

\bibitem[\protect\citeauthoryear{{Cattaneo}, {Blaizot}, {Devriendt} \&
  {Guiderdoni}}{{Cattaneo} et~al.}{2005}]{cattaneo05}
{Cattaneo} A.,  {Blaizot} J.,  {Devriendt} J.,    {Guiderdoni} B.,  2005,
  \mnras, 364, 407

\bibitem[\protect\citeauthoryear{{Coil}, {Hennawi}, {Newman}, {Cooper} \&
  {Davis}}{{Coil} et~al.}{2007}]{coil07}
{Coil} A.~L.,  {Hennawi} J.~F.,  {Newman} J.~A.,  {Cooper} M.~C.,    {Davis}
  M.,  2007, \apj, 654, 115

\bibitem[\protect\citeauthoryear{{Colberg} \& {di Matteo}}{{Colberg} \& {di
  Matteo}}{2008}]{colberg08}
{Colberg} J.~M.,  {di Matteo} T.,  2008, \mnras, 387, 1163

\bibitem[\protect\citeauthoryear{{Cole} \& {Kaiser}}{{Cole} \&
  {Kaiser}}{1989}]{cole89}
{Cole} S.,  {Kaiser} N.,  1989, \mnras, 237, 1127

\bibitem[\protect\citeauthoryear{{Cowie}, {Barger}, {Bautz}, {Brandt} \&
  {Garmire}}{{Cowie} et~al.}{2003}]{cowie03}
{Cowie} L.~L.,  {Barger} A.~J.,  {Bautz} M.~W.,  {Brandt} W.~N.,    {Garmire}
  G.~P.,  2003, \apjl, 584, L57

\bibitem[\protect\citeauthoryear{{Croom}, {Boyle}, {Shanks}, {Smith}, {Miller},
  {Outram}, {Loaring}, {Hoyle} \& {da {\^A}ngela}}{{Croom}
  et~al.}{2005}]{croom05}
{Croom} S.~M.,  {Boyle} B.~J.,  {Shanks} T.,  {Smith} R.~J.,  {Miller} L.,
  {Outram} P.~J.,  {Loaring} N.~S.,  {Hoyle} F.,    {da {\^A}ngela} J.,  2005,
  \mnras, 356, 415

\bibitem[\protect\citeauthoryear{{Croom}, {Smith}, {Boyle}, {Shanks}, {Miller},
  {Outram} \& {Loaring}}{{Croom} et~al.}{2004}]{croom04}
{Croom} S.~M.,  {Smith} R.~J.,  {Boyle} B.~J.,  {Shanks} T.,  {Miller} L.,
  {Outram} P.~J.,    {Loaring} N.~S.,  2004, \mnras, 349, 1397

\bibitem[\protect\citeauthoryear{{Croton}}{{Croton}}{2006}]{croton06b}
{Croton} D.~J.,  2006, \mnras, 369, 1808

\bibitem[\protect\citeauthoryear{{Croton}, {Springel}, {White}, {De Lucia},
  {Frenk}, {Gao}, {Jenkins}, {Kauffmann} et~al.,}{{Croton}
  et~al.}{2006}]{croton06a}
{Croton} D.~J.,  {Springel} V.,  {White} S.~D.~M.,  {De Lucia} G.,  {Frenk}
  C.~S.,  {Gao} L.,  {Jenkins} A.,  {Kauffmann} G.,    et~al., 2006, \mnras,
  365, 11

\bibitem[\protect\citeauthoryear{{da {\^A}ngela}, {Shanks}, {Croom},
  {Weilbacher}, {Brunner}, {Couch}, {Miller}, {Myers} et~al.,}{{da {\^A}ngela}
  et~al.}{2008}]{daAngela08}
{da {\^A}ngela} J.,  {Shanks} T.,  {Croom} S.~M.,  {Weilbacher} P.,  {Brunner}
  R.~J.,  {Couch} W.~J.,  {Miller} L.,  {Myers} A.~D.,    et~al., 2008, \mnras,
  383, 565

\bibitem[\protect\citeauthoryear{{De Lucia} \& {Blaizot}}{{De Lucia} \&
  {Blaizot}}{2007}]{delucia07}
{De Lucia} G.,  {Blaizot} J.,  2007, \mnras, 375, 2

\bibitem[\protect\citeauthoryear{{Di Matteo}, {Springel} \& {Hernquist}}{{Di
  Matteo} et~al.}{2005}]{diMatteo05}
{Di Matteo} T.,  {Springel} V.,    {Hernquist} L.,  2005, \nat, 433, 604

\bibitem[\protect\citeauthoryear{{Elvis}, {Risaliti} \& {Zamorani}}{{Elvis}
  et~al.}{2002}]{elvis02}
{Elvis} M.,  {Risaliti} G.,    {Zamorani} G.,  2002, \apjl, 565, L75

\bibitem[\protect\citeauthoryear{{Fan}, {Narayanan}, {Lupton}, {Strauss},
  {Knapp}, {Becker}, {White}, {Pentericci}, {Leggett} et~al.,}{{Fan}
  et~al.}{2001}]{fan01}
{Fan} X.,  {Narayanan} V.~K.,  {Lupton} R.~H.,  {Strauss} M.~A.,  {Knapp}
  G.~R.,  {Becker} R.~H.,  {White} R.~L.,  {Pentericci} L.,  {Leggett} S.~K.,
   et~al., 2001, \aj, 122, 2833

\bibitem[\protect\citeauthoryear{{Fan}, {White}, {Davis}, {Becker}, {Strauss},
  {Haiman}, {Schneider}, {Gregg} et~al.,}{{Fan} et~al.}{2000}]{fan00}
{Fan} X.,  {White} R.~L.,  {Davis} M.,  {Becker} R.~H.,  {Strauss} M.~A.,
  {Haiman} Z.,  {Schneider} D.~P.,  {Gregg} M.~D.,    et~al., 2000, \aj, 120,
  1167

\bibitem[\protect\citeauthoryear{{Ferrarese}}{{Ferrarese}}{2002}]{ferrarese02}
{Ferrarese} L.,  2002, \apj, 578, 90

\bibitem[\protect\citeauthoryear{{Ferrarese} \& {Ford}}{{Ferrarese} \&
  {Ford}}{2005}]{ferrarese05}
{Ferrarese} L.,  {Ford} H.,  2005, Space Science Reviews, 116, 523

\bibitem[\protect\citeauthoryear{{Ferrarese} \& {Merritt}}{{Ferrarese} \&
  {Merritt}}{2000}]{ferrarese00}
{Ferrarese} L.,  {Merritt} D.,  2000, \apjl, 539, L9

\bibitem[\protect\citeauthoryear{{Fine}, {Croom}, {Miller}, {Babic}, {Moore},
  {Brewer}, {Sharp}, {Boyle} et~al.,}{{Fine} et~al.}{2006}]{fine06}
{Fine} S.,  {Croom} S.~M.,  {Miller} L.,  {Babic} A.,  {Moore} D.,  {Brewer}
  B.,  {Sharp} R.~G.,  {Boyle} B.~J.,    et~al., 2006, \mnras, 373, 613

\bibitem[\protect\citeauthoryear{{Graham} \& {Driver}}{{Graham} \&
  {Driver}}{2007}]{graham07}
{Graham} A.~W.,  {Driver} S.~P.,  2007, \mnras, 380, L15

\bibitem[\protect\citeauthoryear{{Grazian}, {Negrello}, {Moscardini},
  {Cristiani}, {Haehnelt}, {Matarrese}, {Omizzolo} \& {Vanzella}}{{Grazian}
  et~al.}{2004}]{grazian04}
{Grazian} A.,  {Negrello} M.,  {Moscardini} L.,  {Cristiani} S.,  {Haehnelt}
  M.~G.,  {Matarrese} S.,  {Omizzolo} A.,    {Vanzella} E.,  2004, \aj, 127,
  592

\bibitem[\protect\citeauthoryear{{Haehnelt}, {Natarajan} \& {Rees}}{{Haehnelt}
  et~al.}{1998}]{haehnelt98}
{Haehnelt} M.~G.,  {Natarajan} P.,    {Rees} M.~J.,  1998, \mnras, 300, 817

\bibitem[\protect\citeauthoryear{{Haiman} \& {Hui}}{{Haiman} \&
  {Hui}}{2001}]{haiman01}
{Haiman} Z.,  {Hui} L.,  2001, \apj, 547, 27

\bibitem[\protect\citeauthoryear{{H{\"a}ring} \& {Rix}}{{H{\"a}ring} \&
  {Rix}}{2004}]{haring04}
{H{\"a}ring} N.,  {Rix} H.-W.,  2004, \apjl, 604, L89

\bibitem[\protect\citeauthoryear{{Hasinger}, {Miyaji} \& {Schmidt}}{{Hasinger}
  et~al.}{2005}]{hasinger05}
{Hasinger} G.,  {Miyaji} T.,    {Schmidt} M.,  2005, \aap, 441, 417

\bibitem[\protect\citeauthoryear{{Heckman}, {Kauffmann}, {Brinchmann},
  {Charlot}, {Tremonti} \& {White}}{{Heckman} et~al.}{2004}]{heckman04}
{Heckman} T.~M.,  {Kauffmann} G.,  {Brinchmann} J.,  {Charlot} S.,  {Tremonti}
  C.,    {White} S.~D.~M.,  2004, \apj, 613, 109

\bibitem[\protect\citeauthoryear{{Heger} \& {Woosley}}{{Heger} \&
  {Woosley}}{2002}]{heger02}
{Heger} A.,  {Woosley} S.~E.,  2002, \apj, 567, 532

\bibitem[\protect\citeauthoryear{{Hopkins} \& {Hernquist}}{{Hopkins} \&
  {Hernquist}}{2008}]{hopkins08}
{Hopkins} P.~F.,  {Hernquist} L.,  2008, arXiv:0809.3789

\bibitem[\protect\citeauthoryear{{Hopkins}, {Hernquist}, {Martini}, {Cox},
  {Robertson}, {Di Matteo} \& {Springel}}{{Hopkins} et~al.}{2005}]{hopkins05}
{Hopkins} P.~F.,  {Hernquist} L.,  {Martini} P.,  {Cox} T.~J.,  {Robertson} B.,
   {Di Matteo} T.,    {Springel} V.,  2005, \apjl, 625, L71

\bibitem[\protect\citeauthoryear{{Hopkins}, {Lidz}, {Hernquist}, {Coil},
  {Myers}, {Cox} \& {Spergel}}{{Hopkins} et~al.}{2007}]{hopkins07b}
{Hopkins} P.~F.,  {Lidz} A.,  {Hernquist} L.,  {Coil} A.~L.,  {Myers} A.~D.,
  {Cox} T.~J.,    {Spergel} D.~N.,  2007, \apj, 662, 110

\bibitem[\protect\citeauthoryear{{Hopkins}, {Richards} \&
  {Hernquist}}{{Hopkins} et~al.}{2007}]{hopkins07a}
{Hopkins} P.~F.,  {Richards} G.~T.,    {Hernquist} L.,  2007, \apj, 654, 731

\bibitem[\protect\citeauthoryear{{Kauffmann} \& {Haehnelt}}{{Kauffmann} \&
  {Haehnelt}}{2000}]{kauffmann00}
{Kauffmann} G.,  {Haehnelt} M.,  2000, \mnras, 311, 576

\bibitem[\protect\citeauthoryear{{Kirkman} \& {Tytler}}{{Kirkman} \&
  {Tytler}}{2008}]{kirkman08}
{Kirkman} D.,  {Tytler} D.,  2008, \mnras, 391, 1457

\bibitem[\protect\citeauthoryear{{Kollmeier}, {Onken}, {Kochanek}, {Gould},
  {Weinberg}, {Dietrich}, {Cool}, {Dey} et~al.,}{{Kollmeier}
  et~al.}{2006}]{kollmeier06}
{Kollmeier} J.~A.,  {Onken} C.~A.,  {Kochanek} C.~S.,  {Gould} A.,  {Weinberg}
  D.~H.,  {Dietrich} M.,  {Cool} R.,  {Dey} A.,    et~al., 2006, \apj, 648, 128

\bibitem[\protect\citeauthoryear{{Koushiappas}, {Bullock} \&
  {Dekel}}{{Koushiappas} et~al.}{2004}]{koushiappas04}
{Koushiappas} S.~M.,  {Bullock} J.~S.,    {Dekel} A.,  2004, \mnras, 354, 292

\bibitem[\protect\citeauthoryear{{Koushiappas} \& {Zentner}}{{Koushiappas} \&
  {Zentner}}{2006}]{koushiappas06}
{Koushiappas} S.~M.,  {Zentner} A.~R.,  2006, \apj, 639, 7

\bibitem[\protect\citeauthoryear{{Li}, {Kauffmann}, {Wang}, {White}, {Heckman}
  \& {Jing}}{{Li} et~al.}{2006}]{li06}
{Li} C.,  {Kauffmann} G.,  {Wang} L.,  {White} S.~D.~M.,  {Heckman} T.~M.,
  {Jing} Y.~P.,  2006, \mnras, 373, 457

\bibitem[\protect\citeauthoryear{{Lidz}, {Hopkins}, {Cox}, {Hernquist} \&
  {Robertson}}{{Lidz} et~al.}{2006}]{lidz06}
{Lidz} A.,  {Hopkins} P.~F.,  {Cox} T.~J.,  {Hernquist} L.,    {Robertson} B.,
  2006, \apj, 641, 41

\bibitem[\protect\citeauthoryear{{Loeb} \& {Rasio}}{{Loeb} \&
  {Rasio}}{1994}]{loeb94}
{Loeb} A.,  {Rasio} F.~A.,  1994, \apj, 432, 52

\bibitem[\protect\citeauthoryear{{Madau} \& {Rees}}{{Madau} \&
  {Rees}}{2001}]{madau01}
{Madau} P.,  {Rees} M.~J.,  2001, \apjl, 551, L27

\bibitem[\protect\citeauthoryear{{Magorrian}, {Tremaine}, {Richstone},
  {Bender}, {Bower}, {Dressler}, {Faber}, {Gebhardt} et~al.,}{{Magorrian}
  et~al.}{1998}]{magorrian98}
{Magorrian} J.,  {Tremaine} S.,  {Richstone} D.,  {Bender} R.,  {Bower} G.,
  {Dressler} A.,  {Faber} S.~M.,  {Gebhardt} K.,    et~al., 1998, \aj, 115,
  2285

\bibitem[\protect\citeauthoryear{{Malbon}, {Baugh}, {Frenk} \&
  {Lacey}}{{Malbon} et~al.}{2007}]{malbon07}
{Malbon} R.~K.,  {Baugh} C.~M.,  {Frenk} C.~S.,    {Lacey} C.~G.,  2007,
  \mnras, 382, 1394

\bibitem[\protect\citeauthoryear{{Marconi} \& {Hunt}}{{Marconi} \&
  {Hunt}}{2003}]{marconi03}
{Marconi} A.,  {Hunt} L.~K.,  2003, \apjl, 589, L21

\bibitem[\protect\citeauthoryear{{Marconi}, {Risaliti}, {Gilli}, {Hunt},
  {Maiolino} \& {Salvati}}{{Marconi} et~al.}{2004}]{marconi04}
{Marconi} A.,  {Risaliti} G.,  {Gilli} R.,  {Hunt} L.~K.,  {Maiolino} R.,
  {Salvati} M.,  2004, \mnras, 351, 169

\bibitem[\protect\citeauthoryear{{Martini}}{{Martini}}{2004}]{martini04}
{Martini} P.,  2004, in {Ho} L.~C.,  ed., Coevolution of Black Holes and
  Galaxies {QSO Lifetimes}.
pp 169--+

\bibitem[\protect\citeauthoryear{{Martini} \& {Weinberg}}{{Martini} \&
  {Weinberg}}{2001}]{martini01}
{Martini} P.,  {Weinberg} D.~H.,  2001, \apj, 547, 12

\bibitem[\protect\citeauthoryear{{Marulli}, {Bonoli}, {Branchini}, {Gilli},
  {Moscardini} \& {Springel}}{{Marulli} et~al.}{2009}]{marulli09}
{Marulli} F.,  {Bonoli} S.,  {Branchini} E.,  {Gilli} R.,  {Moscardini} L.,
  {Springel} V.,  2009, arXiv:0809.3789

\bibitem[\protect\citeauthoryear{{Marulli}, {Bonoli}, {Branchini}, {Moscardini}
  \& {Springel}}{{Marulli} et~al.}{2008}]{marulli08}
{Marulli} F.,  {Bonoli} S.,  {Branchini} E.,  {Moscardini} L.,    {Springel}
  V.,  2008, \mnras, 385, 1846

\bibitem[\protect\citeauthoryear{{Marulli}, {Branchini}, {Moscardini} \&
  {Volonteri}}{{Marulli} et~al.}{2007}]{marulli07}
{Marulli} F.,  {Branchini} E.,  {Moscardini} L.,    {Volonteri} M.,  2007,
  \mnras, 375, 649

\bibitem[\protect\citeauthoryear{{Marulli}, {Crociani}, {Volonteri},
  {Branchini} \& {Moscardini}}{{Marulli} et~al.}{2006}]{marulli06}
{Marulli} F.,  {Crociani} D.,  {Volonteri} M.,  {Branchini} E.,    {Moscardini}
  L.,  2006, \mnras, 368, 1269

\bibitem[\protect\citeauthoryear{{Merloni} \& {Heinz}}{{Merloni} \&
  {Heinz}}{2008}]{merloni08}
{Merloni} A.,  {Heinz} S.,  2008, \mnras, 388, 1011

\bibitem[\protect\citeauthoryear{{Merloni}, {Rudnick} \& {Di Matteo}}{{Merloni}
  et~al.}{2004}]{merloni04}
{Merloni} A.,  {Rudnick} G.,    {Di Matteo} T.,  2004, \mnras, 354, L37

\bibitem[\protect\citeauthoryear{{Mo} \& {White}}{{Mo} \& {White}}{1996}]{mo96}
{Mo} H.~J.,  {White} S.~D.~M.,  1996, \mnras, 282, 347

\bibitem[\protect\citeauthoryear{{Monaco}, {Fontanot} \& {Taffoni}}{{Monaco}
  et~al.}{2007}]{monaco07}
{Monaco} P.,  {Fontanot} F.,    {Taffoni} G.,  2007, \mnras, 375, 1189

\bibitem[\protect\citeauthoryear{{Myers}, {Brunner}, {Nichol}, {Richards},
  {Schneider} \& {Bahcall}}{{Myers} et~al.}{2007}]{myers07}
{Myers} A.~D.,  {Brunner} R.~J.,  {Nichol} R.~C.,  {Richards} G.~T.,
  {Schneider} D.~P.,    {Bahcall} N.~A.,  2007, \apj, 658, 85

\bibitem[\protect\citeauthoryear{{Netzer} \& {Trakhtenbrot}}{{Netzer} \&
  {Trakhtenbrot}}{2007}]{netzer07}
{Netzer} H.,  {Trakhtenbrot} B.,  2007, \apj, 654, 754

\bibitem[\protect\citeauthoryear{{Padmanabhan}, {White}, {Norberg} \&
  {Porciani}}{{Padmanabhan} et~al.}{2008}]{padmanabhan08}
{Padmanabhan} N.,  {White} M.,  {Norberg} P.,    {Porciani} C.,  2008,
  arXiv:0802.2105

\bibitem[\protect\citeauthoryear{{Peacock}}{{Peacock}}{1999}]{peacock99}
{Peacock} J.~A.,  1999, {Cosmological Physics}.
Cosmological Physics, by John A.~Peacock, pp.~704.~ISBN 052141072X.~Cambridge,
  UK: Cambridge University Press, January 1999.

\bibitem[\protect\citeauthoryear{{Porciani}, {Magliocchetti} \&
  {Norberg}}{{Porciani} et~al.}{2004}]{porciani04}
{Porciani} C.,  {Magliocchetti} M.,    {Norberg} P.,  2004, \mnras, 355, 1010

\bibitem[\protect\citeauthoryear{{Ricotti}, {Ostriker} \& {Gnedin}}{{Ricotti}
  et~al.}{2005}]{ricotti05}
{Ricotti} M.,  {Ostriker} J.~P.,    {Gnedin} N.~Y.,  2005, \mnras, 357, 207

\bibitem[\protect\citeauthoryear{{Ripamonti}, {Mapelli} \&
  {Zaroubi}}{{Ripamonti} et~al.}{2008}]{ripamonti08}
{Ripamonti} E.,  {Mapelli} M.,    {Zaroubi} S.,  2008, \mnras, 387, 158

\bibitem[\protect\citeauthoryear{{Ross}, {Shen}, {Strauss}, {Vanden Berk},
  {Connolly}, {Richards}, {Schneider}, {Weinberg} et~al.,}{{Ross}
  et~al.}{2009}]{ross09}
{Ross} N.~P.,  {Shen} Y.,  {Strauss} M.~A.,  {Vanden Berk} D.~E.,  {Connolly}
  A.~J.,  {Richards} G.~T.,  {Schneider} D.~P.,  {Weinberg} D.~H.,    et~al.,
  2009, ArXiv:0903.3230

\bibitem[\protect\citeauthoryear{{Sesana}, {Haardt}, {Madau} \&
  {Volonteri}}{{Sesana} et~al.}{2005}]{sesana05}
{Sesana} A.,  {Haardt} F.,  {Madau} P.,    {Volonteri} M.,  2005, \apj, 623, 23

\bibitem[\protect\citeauthoryear{{Shakura} \& {Sunyaev}}{{Shakura} \&
  {Sunyaev}}{1973}]{shakura73}
{Shakura} N.~I.,  {Sunyaev} R.~A.,  1973, \aap, 24, 337

\bibitem[\protect\citeauthoryear{{Shankar}, {Lapi}, {Salucci}, {De Zotti} \&
  {Danese}}{{Shankar} et~al.}{2006}]{shankar06}
{Shankar} F.,  {Lapi} A.,  {Salucci} P.,  {De Zotti} G.,    {Danese} L.,  2006,
  \apj, 643, 14

\bibitem[\protect\citeauthoryear{{Shankar}, {Salucci}, {Granato}, {De Zotti} \&
  {Danese}}{{Shankar} et~al.}{2004}]{shankar04}
{Shankar} F.,  {Salucci} P.,  {Granato} G.~L.,  {De Zotti} G.,    {Danese} L.,
  2004, \mnras, 354, 1020

\bibitem[\protect\citeauthoryear{{Shen}, {Greene}, {Strauss}, {Richards} \&
  {Schneider}}{{Shen} et~al.}{2008}]{shen08a}
{Shen} Y.,  {Greene} J.~E.,  {Strauss} M.~A.,  {Richards} G.~T.,    {Schneider}
  D.~P.,  2008, \apj, 680, 169

\bibitem[\protect\citeauthoryear{{Shen}, {Strauss}, {Oguri}, {Hennawi}, {Fan},
  {Richards}, {Hall}, {Gunn} et~al.,}{{Shen} et~al.}{2007}]{shen07}
{Shen} Y.,  {Strauss} M.~A.,  {Oguri} M.,  {Hennawi} J.~F.,  {Fan} X.,
  {Richards} G.~T.,  {Hall} P.~B.,  {Gunn} J.~E.,    et~al., 2007, \aj, 133,
  2222

\bibitem[\protect\citeauthoryear{{Shen}, {Strauss}, {Ross}, {Hall}, {Lin},
  {Richards}, {Schneider}, {Weinberg} et~al.,}{{Shen} et~al.}{2008}]{shen08b}
{Shen} Y.,  {Strauss} M.~A.,  {Ross} N.~P.,  {Hall} P.~B.,  {Lin} Y.-T.,
  {Richards} G.~T.,  {Schneider} D.~P.,  {Weinberg} D.~H.,    et~al., 2008,
  ArXiv:0810.4144

\bibitem[\protect\citeauthoryear{{Sheth} \& {Tormen}}{{Sheth} \&
  {Tormen}}{1999}]{sheth99}
{Sheth} R.~K.,  {Tormen} G.,  1999, \mnras, 308, 119

\bibitem[\protect\citeauthoryear{{Silk} \& {Rees}}{{Silk} \&
  {Rees}}{1998}]{silk98}
{Silk} J.,  {Rees} M.~J.,  1998, \aap, 331, L1

\bibitem[\protect\citeauthoryear{{Soltan}}{{Soltan}}{1982}]{soltan82}
{Soltan} A.,  1982, \mnras, 200, 115

\bibitem[\protect\citeauthoryear{{Spergel}, {Verde}, {Peiris}, {Komatsu},
  {Nolta}, {Bennett}, {Halpern}, {Hinshaw} et~al.,}{{Spergel}
  et~al.}{2003}]{spergel03}
{Spergel} D.~N.,  {Verde} L.,  {Peiris} H.~V.,  {Komatsu} E.,  {Nolta} M.~R.,
  {Bennett} C.~L.,  {Halpern} M.,  {Hinshaw} G.,    et~al., 2003, \apjs, 148,
  175

\bibitem[\protect\citeauthoryear{{Springel}, {Di Matteo} \&
  {Hernquist}}{{Springel} et~al.}{2005}]{springel05a}
{Springel} V.,  {Di Matteo} T.,    {Hernquist} L.,  2005, \mnras, 361, 776

\bibitem[\protect\citeauthoryear{{Springel}, {White}, {Jenkins}, {Frenk},
  {Yoshida}, {Gao}, {Navarro}, {Thacker} et~al.,}{{Springel}
  et~al.}{2005}]{springel05b}
{Springel} V.,  {White} S.~D.~M.,  {Jenkins} A.,  {Frenk} C.~S.,  {Yoshida} N.,
   {Gao} L.,  {Navarro} J.,  {Thacker} R.,    et~al., 2005, \nat, 435, 629

\bibitem[\protect\citeauthoryear{{Springel}, {White}, {Tormen} \&
  {Kauffmann}}{{Springel} et~al.}{2001}]{springel01}
{Springel} V.,  {White} S.~D.~M.,  {Tormen} G.,    {Kauffmann} G.,  2001,
  \mnras, 328, 726

\bibitem[\protect\citeauthoryear{{Steffen}, {Barger}, {Cowie}, {Mushotzky} \&
  {Yang}}{{Steffen} et~al.}{2003}]{steffen03}
{Steffen} A.~T.,  {Barger} A.~J.,  {Cowie} L.~L.,  {Mushotzky} R.~F.,    {Yang}
  Y.,  2003, \apjl, 596, L23

\bibitem[\protect\citeauthoryear{{Thacker}, {Scannapieco}, {Couchman} \&
  {Richardson}}{{Thacker} et~al.}{2009}]{thacker08}
{Thacker} R.~J.,  {Scannapieco} E.,  {Couchman} H.~M.~P.,    {Richardson} M.,
  2009, \apj, 693, 552

\bibitem[\protect\citeauthoryear{{Tremaine}, {Gebhardt}, {Bender}, {Bower},
  {Dressler}, {Faber}, {Filippenko}, {Green} et~al.,}{{Tremaine}
  et~al.}{2002}]{tremaine02}
{Tremaine} S.,  {Gebhardt} K.,  {Bender} R.,  {Bower} G.,  {Dressler} A.,
  {Faber} S.~M.,  {Filippenko} A.~V.,  {Green} R.,    et~al., 2002, \apj, 574,
  740

\bibitem[\protect\citeauthoryear{{Ueda}, {Akiyama}, {Ohta} \& {Miyaji}}{{Ueda}
  et~al.}{2003}]{ueda03}
{Ueda} Y.,  {Akiyama} M.,  {Ohta} K.,    {Miyaji} T.,  2003, \apj, 598, 886

\bibitem[\protect\citeauthoryear{{White}, {Martini} \& {Cohn}}{{White}
  et~al.}{2008}]{white07}
{White} M.,  {Martini} P.,    {Cohn} J.~D.,  2008, \mnras, 390, 1179

\bibitem[\protect\citeauthoryear{{York}, {Adelman}, {Anderson} Jr., {Anderson},
  {Annis}, {Bahcall}, {Bakken}, {Barkhouser} et~al.,}{{York}
  et~al.}{2000}]{york00}
{York} D.~G.,  {Adelman} J.,  {Anderson} Jr. J.~E.,  {Anderson} S.~F.,  {Annis}
  J.,  {Bahcall} N.~A.,  {Bakken} J.~A.,  {Barkhouser} R.,    et~al., 2000,
  \aj, 120, 1579

\bibitem[\protect\citeauthoryear{{Yu} \& {Tremaine}}{{Yu} \&
  {Tremaine}}{2002}]{yu02}
{Yu} Q.,  {Tremaine} S.,  2002, \mnras, 335, 965

\end{thebibliography}

\label{lastpage}

\end{document}